\documentclass[prd,showpacs,showkeys,floatfix,nofootinbib,eqsecnum,
               preprint,12pt,tightenlines,fleqn]{revtex4} 

\usepackage{amsmath,amssymb,revsymb,graphicx,dcolumn}

\newcommand{\beq}{\begin{equation}}
\newcommand{\eeq}{\end{equation}}
\newcommand{\beqa}{\begin{eqnarray}}
\newcommand{\eeqa}{\end{eqnarray}}
\newcommand{\bsubeqs}{\begin{subequations}}
\newcommand{\esubeqs}{\end{subequations}}

\newcommand{\epsilondimless}{e}    

\begin{document}
%
%
\noindent Phys. Rev. D 85, 103508 (2012) \hfill arXiv:1109.4915\newline\vspace*{2mm}
%
\vspace*{8mm}\newline
\title[MAIN COSMOLOGICAL CONSTANT PROBLEM]
      {Possible solution to the main cosmological constant problem\vspace*{5mm}}
\author{V. Emelyanov}
\email{slawa@particle.uni-karlsruhe.de}
\author{F.R. Klinkhamer}
\email{frans.klinkhamer@kit.edu}\affiliation{ \mbox{Institute for
Theoretical Physics, University of Karlsruhe,} Karlsruhe Institute of
Technology, 76128 Karlsruhe, Germany\\}

\begin{abstract}
\vspace*{2.5mm}\noindent
A modified-gravity-type model of two hypothetical massless
vector fields is presented. These vector fields are gravitationally
coupled to standard matter and an effective cosmological constant.
Considered in a cosmological context, the vector fields dynamically
cancel an arbitrary cosmological constant,
and flat Minkowski spacetime appears as the limit of
attractor-type solutions of the field equations.
Asymptotically, the field equations give rise
to a standard Friedmann-Robertson-Walker universe
and standard Newtonian gravitational dynamics of small systems.
\end{abstract}



\maketitle

\section{Introduction and summary}
\label{sec:intro}

The main cosmological constant problem (CCP No. 1 or CCP1, for short) lies in
the apparent conflict between certain theoretical expectations and
experimental facts (see, e.g., Ref.~\cite{Weinberg1989}
for an extensive review). The
key theoretical expectation is that the zero-point energy of the quantum
fields in the equilibrium vacuum state naturally produces an unsuppressed
effective cosmological constant $\Lambda$ in the classical gravitational
field equation. The key experimental fact is the observed negligible value
of $\Lambda$. The qualifications \mbox{`unsuppressed'} and `negligible'
refer to the known energy scales of elementary particle physics.
Hence, CCP1 motivates
us to discover the mechanism which cancels the gravitational effects
of this zero-point energy, without fine-tuning the theory.

Several years ago, Dolgov~\cite{Dolgov1985-1997} proposed
a remarkable solution to CCP1 by having
an evolving massless vector field which
dynamically cancels the effective cosmological constant $\Lambda$.
The original Dolgov model, however, runs into two obstacles.
The first obstacle~\cite{RubakovTinyakov1999} is that
the steadily increasing vector field ruins the Newtonian
gravitational dynamics of a localized matter distribution, e.g.,
the matter of the Solar System. The second obstacle, already noted by
Dolgov himself, is that the expansion of the asymptotic Universe
is too fast, inexorably diluting any standard-matter component
initially present.

Inspired by the $q$--theory
approach~\cite{KV2008-2010,KV2011-review} to CCP1,
two extended vector-field models have been constructed, which
circumvent each obstacle
separately~\cite{EmelyanovKlinkhamer2011-CCP1-NEWTON,
EmelyanovKlinkhamer2011-CCP1-FRW}.
The question is whether or not there exists a further
extended vector-field model which deals with
both obstacles simultaneously. The present article
answers this question affirmatively by doing the obvious,
namely, by combining the two previous extended models.

With this final extended vector-field model, we have a possible
solution of the main cosmological constant problem
and no unwanted side effects. Indeed, the final extended vector-field model
cancels an arbitrary (Planck-scale) cosmological constant $\Lambda$
without fine-tuning, while maintaining
the standard local Newtonian gravitational dynamics
and providing for an acceptable late-universe Hubble expansion
(there may still be an inflationary phase in
the very early universe~\cite{Klinkhamer2011-CCP1-inflation}).
But, this is only a `possible' solution, because it is not
clear if such massless vector fields exist in reality
(having a consistent quantum theory and a mechanism
to guarantee their masslessness).
A further \textit{caveat} on this `possible' solution is mentioned
in Endnote~\cite{Endnote-graviton-mass} which is called in
Sec.~\ref{Standard-local-Newtonian-dynamics}.

Taking for granted that CCP1 has been solved in principle,
the next problem (CCP2) is to explain the small but nonzero value
measured in the actual nonequilibrium Universe. This problem
lies outside the scope of the present article.
Some relevant remarks can be found
in the recent review~\cite{KV2011-review}, which contains, moreover,
a brief summary of $q$--theory.

The outline of this article is straightforward: first,
the model is defined (Sec.~\ref{sec:two-vector-field-model}), then,
the homogeneous background solution is determined
(Sec.~\ref{sec:Asymptotic-solution})
and found to correspond to an attractor-type solution
(Sec.~\ref{sec:Attractor-type-solutions}),
and, finally, the local gravitational dynamics of small-scale systems
is shown to be Newtonian (Sec.~\ref{sec:Second-order-perturbations}).
Two appendices give mathematical proofs for the existence
of attractor solutions in related but simpler vector-field models,
namely, the original Dolgov model~\cite{Dolgov1985-1997} and the
model of our second article~\cite{EmelyanovKlinkhamer2011-CCP1-FRW}.

All calculations contained in this article are analytic.
The main results are, first,
the exact solution \eqref{eq:A0-B0-H-rM-sol}
with constants \eqref{eq:values-sol} and
$\Lambda$--cancellation \eqref{eq:Lambda-nullification}
and, second, the vanishing modification \eqref{eq:ThetaIsZero}
of the weak-field gravity theory \eqref{eq:weak-grav-eq-h-S-T}
for a localized matter distribution in a perfect-equilibrium background.

\section{Two-vector-field model}
\label{sec:two-vector-field-model}

The model is presented in Sec.~\ref{subsec:action},
together with appropriate cosmological \textit{Ans\"{a}tze}
for the fields. The reduced field equations are given
in Secs.~\ref{subsec:Reduced-vector-field-equations}
and \ref{sec:Generalized-FRW-equations}.

\subsection{Action and Ans\"{a}tze}\label{subsec:action}

Consider a model of two massless vector fields,
$A_{\alpha}(x)$ and $B_{\alpha}(x)$. This model is governed by
the following effective action ($\hbar=c=1$):
\bsubeqs\label{eq:model-action-EPlanck}
\beqa\label{eq:model-action}
&&
S_\text{eff}[g,\, A,\, B,\,\phi]=
\nonumber\\[1mm]&&
-\int\,d^4x\,\sqrt{-\text{det}(g)}\;
\bigg( \frac{1}{2}\,(E_\text{Planck})^2\,R[g]
+ \epsilon(F_{A},F_{B}) + \Lambda
+ \mathcal{L}_{M}[g,\,\phi]\bigg)\,,
\\[2mm]\label{eq:EPlanck}
&&
E_\text{Planck} \equiv (8\pi G)^{-1/2}\,,
\eeqa
\esubeqs
with the Ricci scalar $R(x)$ of the metric $g_{\alpha\beta}(x)$,
the effective cosmological constant $\Lambda$,
a generic massless matter field $\phi(x)$
with a standard Lagrange density $\mathcal{L}_{M}(x)$,
and a function $\epsilon(F_{A},F_{B})$ of the following two
auxiliary variables $F_{A}(x)$ and $F_{B}(x)$:%
\bsubeqs\label{eq:def-FA-Q3A-FB-Q3B}
\beqa\label{eq:def-FA-Q3A}
F_{A}[g,\, A] &\equiv& (Q_{3A})^2 - \frac{1}{2}\,R\,A_{\alpha}\,A^{\alpha}\,,
\quad
Q_{3A}[g,\, A] \equiv \nabla^{\alpha}A_{\alpha}\,,
\\[2mm]
\label{eq:def-FB-Q3B}
F_{B}[g,\, B] &\equiv& (Q_{3B})^2 - \frac{1}{2}\,R\,B_{\alpha}\,B^{\alpha}\,,
\quad
Q_{3B}[g,\, B] \equiv \nabla^{\alpha}B_{\alpha}\,,
\eeqa
\esubeqs
where $\nabla_{\alpha}$ denotes the covariant derivative
(later also written as a semicolon in front of the
relevant spacetime index).
As will become clear at the very end of this article
(Sec.~\ref{Standard-local-Newtonian-dynamics}),
the coupling constant $G$
entering the reduced Planck energy \eqref{eq:EPlanck}
can be identified with Newton's gravitational coupling
constant $G_N$, first measured by Cavendish.
The magnitude of the cosmological constant is considered to be of the
order of the Planck energy, $|\Lambda| \sim (E_\text{Planck})^4$.
For the moment, we set the standard term $\mathcal{L}_{M}$
in \eqref{eq:model-action} to zero
[possible zero-point-energy contributions from the $\phi$ field
have already been included in $\Lambda$,
assuming a proper (relativistic) regularization
tracing back to the fundamental microscopic theory].

Combining the \textit{Ans\"{a}tze} of our previous
work~\cite{EmelyanovKlinkhamer2011-CCP1-NEWTON,EmelyanovKlinkhamer2011-CCP1-FRW},
we take the following special function:
\beqa\label{eq:epsilon-Ansatz}
\epsilon(F_{A},F_{B}) &=& (E_\text{Planck})^4\;
\bigg(a\;\frac{F_{A}}{F_{B}} +b\;\frac{F_{B}}{F_{A}}\bigg)\,,
\eeqa
with numerical constants $a=\pm 1$ and $b=-a$.
In explicit calculations later on, we will use the
values $a=-b=1$. The function \eqref{eq:epsilon-Ansatz} possesses the
following symmetry properties:
\bsubeqs\label{eq:property-1-2}
\beqa\label{eq:property-1}
F_{A}\, \frac{\partial\,\epsilon}{\partial\,F_{A}} +
F_{B}\, \frac{\partial\,\epsilon}{\partial\,F_{B}} &=& 0\,,
\eeqa
\beqa\label{eq:property-2}
F_{A}^2\, \frac{\partial^{\,2}\,\epsilon}{\partial\,F_{A}^2} +
2F_{A}F_{B}\, \frac{\partial^{\,2}\,\epsilon}{\partial\,F_{A}\partial\,F_{B}} +
F_{B}^2\, \frac{\partial^{\,2}\,\epsilon}{\partial\,F_{B}^2} &=& 0\,,
\eeqa
\esubeqs
which will turn out to be crucial for
the preservation of standard Newtonian gravity on small scales
(Sec.~\ref{Standard-local-Newtonian-dynamics}).

In this article, we start by considering a spatially flat,
homogeneous, and isotropic universe.
The corresponding  Robertson--Walker (RW) metric
in suitable spacetime coordinates
is given by:
\beqa\label{eq:RW-metric}
\Big(g_{\alpha\beta}(x_1,\,x_2,\,x_3,\,t)\Big) &=&
\Big(\textrm{diag}\big[1,\,-a^2(t),\,-a^2(t),\,-a^2(t)\big]\Big)\,,
\eeqa
where $a(t)$ is the scale factor as a function of
cosmic time $t$. The usual Hubble parameter is defined
by $H\equiv (d a/d t)/a$.

The following \textit{Ans\"{a}tze}~\cite{Dolgov1985-1997}
for the background vector fields are consistent
with the homogeneous and isotropic background
metric \eqref{eq:RW-metric}:
\bsubeqs\label{eq:Ansatz-A-B}
\beqa\label{eq:Ansatz-A}
A_{\alpha}(x_1,\,x_2,\,x_3,\,t) = A_{0}(t)\;\delta_{\alpha}^0\,,\\[2mm]
\label{eq:Ansatz-B}
B_{\alpha}(x_1,\,x_2,\,x_3,\,t) = B_{0}(t)\;\delta_{\alpha}^0\,,
\eeqa
\esubeqs
involving only two functions of $t$.

\subsection{Reduced vector-field equations}
\label{subsec:Reduced-vector-field-equations}

The variational principle for the
vector fields of action \eqref{eq:model-action}
gives the following two equations:
\bsubeqs\label{eq:vector-field-eq-A-B}
\beqa\label{eq:vector-field-eq-A}
\Big(\nabla_{\alpha}Q_{3A} + \frac{1}{2}\,R\,A_{\alpha}\Big)\,
\frac{\partial\,\epsilon}{\partial\,F_{A}} +
Q_{3A}\,\nabla_{\alpha}
\bigg(\frac{\partial\,\epsilon}{\partial\,F_{A}}\bigg) &=& 0\,,
\\[2mm]\label{eq:vector-field-eq-B}
\Big(\nabla_{\alpha}Q_{3B} + \frac{1}{2}\,R\,B_{\alpha}\Big)\,
\frac{\partial\,\epsilon}{\partial\,F_{B}} +
Q_{3B}\,\nabla_{\alpha}
\bigg(\frac{\partial\,\epsilon}{\partial\,F_{B}}\bigg) &=& 0\,.
\eeqa
\esubeqs

The \textit{Ans\"{a}tze} \eqref{eq:RW-metric}
and \eqref{eq:Ansatz-A-B}
reduce the eight partial differential equations
($\ref{eq:vector-field-eq-A}$) and
($\ref{eq:vector-field-eq-B}$) to the following two
ordinary differential equations (ODEs):
\bsubeqs\label{eq:reduced-field-eq-A-B}
\beqa\label{eq:reduced-field-eq-A}
\Big(\ddot{A}_{0} + 3\,H\,\dot{A}_{0} - 6\,H^2\,A_{0}\Big)\,
\frac{\partial\,\epsilon}{\partial\,F_{A}} +
\Big(\dot{A}_{0} + 3\,H\,A_{0}\Big)\,\frac{d}{dt}
\bigg(\frac{\partial\,\epsilon}{\partial\,F_{A}}\bigg) &=& 0\,,
\\[2mm]\label{eq:reduced-field-eq-B}
\Big(\ddot{B}_{0} + 3\,H\,\dot{B}_{0} - 6\,H^2\,B_{0}\Big)\,
\frac{\partial\,\epsilon}{\partial\,F_{B}} +
\Big(\dot{B}_{0} + 3\,H\,B_{0}\Big)\,\frac{d}{dt}
\bigg(\frac{\partial\,\epsilon}{\partial\,F_{B}}\bigg) &=& 0\,,
\eeqa
\esubeqs
where the overdot stands for differentiation with respect to the
cosmic time $t$. Similarly, the $q$--theory-type
variables from \eqref{eq:def-FA-Q3A-FB-Q3B} become
\bsubeqs\label{eq:FRW-Q3A-Q3B}
\beqa\label{eq:FRW-Q3A}
Q_{3A}(t) &=& \frac{d}{d t}\,A_{0}(t)+3\, H\, A_{0}(t)\,,
\\[2mm]\label{eq:FRW-Q3B}
Q_{3B}(t) &=& \frac{d}{d t}\,B_{0}(t)+3\, H\, B_{0}(t)\,.
\eeqa
\esubeqs
Observe that having $A_{0}\propto t$ and $H\propto 1/t$
makes $Q_{3A}$ in \eqref{eq:FRW-Q3A}
into a genuine (spacetime-independent) $q$--theory
variable~\cite{KV2008-2010} and similarly for $Q_{3B}$
in \eqref{eq:FRW-Q3B}.

\subsection{Generalized FRW equations}
\label{sec:Generalized-FRW-equations}

The energy-momentum tensor of the vector fields is
calculated by varying the action with respect to the metric
tensor $g_{\alpha\beta}$. This tensor is found to be given by
\beqa
T_{\alpha\beta} &=&
\bigg(\epsilon(F_{A},F_{B})
- 2\,F_{A}\,\frac{\partial\,\epsilon}{\partial\,F_{A}}
- 2\,F_{B}\,\frac{\partial\,\epsilon}{\partial\,F_{B}}
\bigg)\,g_{\alpha\beta}
\nonumber\\[2mm]&&
+ \frac{\partial\,\epsilon}{\partial\,F_{A}}\,
\Big(R_{\alpha\beta}\,A^2 - R\,A_{\alpha}A_{\beta}\Big)
+ \frac{\partial\,\epsilon}{\partial\,F_{B}}
\Big(R_{\alpha\beta}\,B^2 - R\,B_{\alpha}B_{\beta}\Big)
\nonumber\\[2mm]&&
-\nabla_{\alpha}\nabla_{\beta}
\bigg(A^2\,\frac{\partial\,\epsilon}{\partial\,F_{A}}
    +B^2\,\frac{\partial\,\epsilon}{\partial\,F_{B}}\bigg)
+g_{\alpha\beta}\,\nabla^2\,
\bigg(A^2\,\frac{\partial\,\epsilon}{\partial\,F_{A}}
   + B^2\,\frac{\partial\,\epsilon}{\partial\,F_{B}}\bigg)\,.
\eeqa
At this time, we also introduce a contribution to the
total energy-momentum tensor from the
standard-matter sector of the theory, that is, we consider
having $\mathcal{L}_{M}\ne 0$ in the original
effective action \eqref{eq:model-action}.
In the cosmological context, the standard-matter component is
described by a homogenous relativistic fluid.
Note that this physical setup is not altogether unrealistic,
as the masses of the standard-model particles
are negligible for temperatures
$T \sim E_\text{Planck} \gg 10^2\;\text{GeV}$.

 From the previous \textit{Ans\"{a}tze} \eqref{eq:epsilon-Ansatz},
\eqref{eq:RW-metric}, and \eqref{eq:Ansatz-A-B},
the generalized Friedmann--Robertson--Walker (FRW) equations
and the standard-matter energy-conservation equation are
\bsubeqs\label{eq:FRW-eqs-Friedmann-Einstein-matter}
\beqa\label{eq:FRW-eqs-Friedmann}
3\, H^2 &=&
(E_\text{Planck})^{-2}\;\Big[\Lambda + \rho(A,B)+\rho_M\Big]\,,
\\[2mm]\label{eq:FRW-eqs-Einstein}
2\, \dot{H} + 3\, H^2 &=&
(E_\text{Planck})^{-2}\;\Big[\Lambda - P(A,B)- w_M\,\rho_M\Big]\,,
\\[2mm]\label{eq:FRW-eqs-matter}
\dot{\rho}_M  &=&-3\,(1+w_M)\,H\,\rho_M\,,
\eeqa
\esubeqs
where the last equation describes the adiabatic evolution of a perfect
relativistic fluid with a homogeneous energy density $\rho_M(t)$
and pressure $P_M(t)=w_M\,\rho_M(t)$
for constant equation-of-state parameter $w_{M}=1/3$.
The vector-field energy density
(from $T_{0}^{\;\;0} = \rho$) and isotropic pressure
(from $T_{j}^{\;\;i} = -P\,\delta_{j}^{\;\;i}$) appearing
in \eqref{eq:FRW-eqs-Friedmann-Einstein-matter} are given by
\bsubeqs\label{eq:FRW-rhoAB-PAB}
\beqa\label{eq:FRW-rhoAB}
\rho(A,B) &=& \epsilon(F_{A},F_{B})
+ 3\,\big(\dot{H} + 3\,H^2\big)\,
\bigg(A_{0}^2\,\frac{\partial\,\epsilon}{\partial\,F_{A}} +
B_{0}^2\,\frac{\partial\,\epsilon}{\partial\,F_{B}}\bigg)
\nonumber\\[1mm]&&
+ 3\,H\frac{d}{dt}
\bigg(A_{0}^2\,\frac{\partial\,\epsilon}{\partial\,F_{A}} +
B_{0}^2\,\frac{\partial\,\epsilon}{\partial\,F_{B}}\bigg)\,,
\\[2mm]\label{eq:FRW-PAB}
P(A,B) &=& - \epsilon(F_{A},F_{B})
+ \big(\dot{H} + 3\,H^2\big)\,
\bigg(A_{0}^2\,\frac{\partial\,\epsilon}{\partial\,F_{A}} +
B_{0}^2\,\frac{\partial\,\epsilon}{\partial\,F_{B}}\bigg)
\nonumber\\[1mm]&&
- 2\,H\frac{d}{dt}
\bigg(A_{0}^2\,\frac{\partial\,\epsilon}{\partial\,F_{A}} +
B_{0}^2\,\frac{\partial\,\epsilon}{\partial\,F_{B}}\bigg)
- \frac{d^2}{dt^2}
\bigg(A_{0}^2\,\frac{\partial\,\epsilon}{\partial\,F_{A}} +
B_{0}^2\,\frac{\partial\,\epsilon}{\partial\,F_{B}}\bigg)\,,
\eeqa
\esubeqs
where the symmetry property ($\ref{eq:property-1}$) has been
taken into account. As \eqref{eq:FRW-rhoAB}
contains a term $\dot{H}$, for example, it is clear that
\eqref{eq:FRW-eqs-Friedmann} is not the standard
Friedmann equation.

\section{Asymptotic solution}
\label{sec:Asymptotic-solution}

It is a straightforward exercise to determine the
asymptotic  ($t\to\infty$) solution from the reduced field equations
as given in Sec.~\ref{sec:two-vector-field-model}.
In a first reading, it is possible to skip the technical details
and to jump ahead to Sec.~\ref{subsec:Lambda-cancellation},
which contains the main physics result of this section.

\subsection{Dimensionless ODEs}
\label{subsec:Dimensionless ODEs}

As in our previous
articles~\cite{EmelyanovKlinkhamer2011-CCP1-NEWTON,
EmelyanovKlinkhamer2011-CCP1-FRW}, we introduce
dimensionless variables by rescaling with appropriate powers
of the reduced Planck energy $E_\text{Planck}$
without additional numerical factors. Specifically, we replace
\bsubeqs\label{eq:dimensionless-variables}
\beqa
\big\{\Lambda,\,\epsilon,\, t,\,  H\big\}
&\to&
\big\{\lambda,\, \epsilondimless,\,  \tau,\, h\big\}\,,\\[2mm]
\big\{Q_{3A},\, Q_{3B},\, A_{0},\, B_{0},\,\rho_{M} \big\}
&\to&
\big\{q_{3A},\, q_{3B},\, v,\, w,\,r_{M} \big\}\,.
\eeqa
\esubeqs

The following dimensionless ODEs for the vector fields $v(\tau)$
and $w(\tau)$
and the Hubble parameter $h(\tau)$ result from the previous vector-field,
generalized Friedmann, and matter energy-conservation equations:%
\bsubeqs\label{eq:ODEs-v-w-h-rM}
\beqa\label{eq:ODEs-v}
0&=&
\Big(\ddot{v} + 3\, h\, \dot{v} -  6\, h^2\, v \Big)\,
\frac{\partial \epsilondimless}{\partial f_A}
+\Big(\dot{v} +  3\, h\, v\Big)\,
\frac{d}{d\tau}\left(\frac{\partial \epsilondimless}{\partial f_A}\right)
\,,\\[2mm]
\label{eq:ODEs-w}
0&=&
\Big(\ddot{w} + 3\, h\, \dot{w} -  6\, h^2\, w \Big)\,
\frac{\partial \epsilondimless}{\partial f_B}
+\Big(\dot{w} +  3\, h\, w\Big)\,
\frac{d}{d\tau}\left(\frac{\partial\epsilondimless}{\partial f_B}\right)
\,,\\[2mm]
\label{eq:ODEs-h}
0 &=&3\, h^2- \lambda - \widetilde{\epsilondimless} - r_{M} -
        3\,\Big(\dot{h} + 3\, h^2\Big)\;g_{AB}
        - 3\, h\, \frac{d}{d\tau}\, g_{AB}\,,\\[2mm]
\label{eq:ODEs-rM}
0 &=& \dot{r}_{M} + 4\,h\,r_{M}\,,
\eeqa
\esubeqs
where  the overdot now
stands for differentiation with respect to the dimensionless
cosmic time $\tau$.
In addition, we have the following definitions:
\bsubeqs\label{eq:def-epsilondimless,fA,fB,gAB}
\beqa
\epsilondimless &=& f_A/f_B-f_B/f_A
\,,\\[2mm]
\widetilde{\epsilondimless} &=&\epsilondimless
-2\, f_A\, \big(\partial\epsilondimless/\partial f_A\big)
-2\, f_B\, \big(\partial\epsilondimless/\partial f_B\big)
= \epsilondimless
\,,\\[2mm]
f_{A} &=&(\dot{v} +  3\, h\, v)^2 +3\,(\dot{h} +2\,h^2)\,v^2
\,,\\[2mm]
f_{B} &=&(\dot{w} +  3\, h\, w)^2 +3\,(\dot{h} +2\,h^2)\,w^2
\,,\\[2mm]
g_{AB} &=&
v^2\,\big(\partial\epsilondimless/\partial f_A\big)+
w^2\,\big(\partial\epsilondimless/\partial f_B\big)\,.
\eeqa
\esubeqs
A further FRW equation, given by the dimensionless version of
\eqref{eq:FRW-eqs-Einstein},
can be shown to be consistent with the above ODEs.

Using the symmetry property \eqref{eq:property-1}
in \eqref{eq:ODEs-v} and  \eqref{eq:ODEs-w},
it can be shown that boundary conditions at $\tau=\tau_{0}$
with $v(\tau_{0})/w(\tau_{0}) = \dot{v}(\tau_{0})/\dot{w}(\tau_{0})$
give proportional $v(\tau)$ and $w(\tau)$ solutions:
$v(\tau)=[v(\tau_{0})/w(\tau_{0})]\,w(\tau)$.

\subsection{Expansion coefficients}
\label{subsec:Expansion-coefficients}

The asymptotic solution of the differential Eqs.~\eqref{eq:ODEs-v-w-h-rM},
for $\lambda$ of arbitrary sign, is given by the following series:
\bsubeqs\label{eq:asymp-sol}
\beqa
v(\tau) &=&
\alpha_{0}\;\tau
+ \alpha_1
+ \alpha_2\;\tau^{-1}
+\text{O}(\tau^{-2})
\,,\\[2mm]
w(\tau) &=&
\beta_{0}\;\tau
+ \beta_1
+ \beta_2\;\tau^{-1}
+\text{O}(\tau^{-2})
\,,\\[2mm]
h(\tau) &=&
\gamma_{0}\;\tau^{-1}
+ \gamma_1\;\tau^{-2}
+ \gamma_2\;\tau^{-3}
+\text{O}(\tau^{-4})
\,,\\[2mm]
r_{M}(\tau) &=&\delta_{0}\;\tau^{-2}
+\delta_1\;\tau^{-3}
+\delta_2\;\tau^{-4}
+\text{O}(\tau^{-5}) \,,
\eeqa
\esubeqs
with leading-order coefficients:
\bsubeqs\label{eq:LO-coeffs}
\beqa
\alpha_{0} &\equiv&  1 \,,
\\[2mm]
\beta_{0} &=&
\pm \,\sqrt{(\lambda/2) + \sqrt{1 +  (\lambda/2)^2}} \,,
\\[2mm]
\gamma_{0} &=&  1/2 \,,
\\[2mm]\label{eq:LO-coeff-delta0}
\delta_{0} &=& 3/4\,,
\eeqa
\esubeqs
next-to-leading-order coefficients:
\bsubeqs\label{eq:NLO-coeffs}
\beqa\label{eq:NLO-coeff-alpha1}
\alpha_1 &=&-2\,\alpha_{0}\,\gamma_1\,,
\\[2mm]\label{eq:NLO-coeff-beta1}
\beta_1 &=&-2\,\beta_{0}\,\gamma_1\,,
\\[2mm]\label{eq:NLO-coeff-gamma1}
\gamma_1 &=& \gamma_1\,,\\[2mm]\label{eq:NLO-coeff-delta1}
\delta_1 &=& 3\,\gamma_1\,,
\eeqa
\esubeqs
and next-to-next-to-leading-order coefficients:
\bsubeqs\label{eq:NNLO-coeffs}
\beqa\label{eq:NNLO-coeff-alpha2}
 \alpha_2 &=& 0\,,
\\[2mm]\label{eq:NNLO-coeff-beta2}
\beta_2 &=& 0\,,
\\[2mm]\label{eq:NNLO-coeff-gamma2}
\gamma_2 &=& 2\,(\gamma_1)^2\,,
\\[2mm]\label{eq:NNLO-coeff-delta2}
\delta_2 &=& 9\,(\gamma_1)^2 \,.
\eeqa
\esubeqs
These vector and metric fields have
only one arbitrary constant, $\gamma_{1}$,
which we interpret as being due to the time-shift
invariance of the equations ($\tau\to\tau+\text{const}$).
The general (attractor-type) solution of the three second-order ODEs
and the single first-order ODE in \eqref{eq:ODEs-v-w-h-rM}
will have seven arbitrary constants
(see Sec.~\ref{subsec:Series-attractor-type-behavior}).

Different starting values of $v(\tau)$, $w(\tau)$, $h(\tau)$,
and $r_{M}(\tau)$, at large enough $\tau=\tau_\text{start}$
and in an appropriate domain,
give different values of $\gamma_{1}$.
Excluded starting values are those with
$\{v(\tau_\text{start}),\,\dot{v}(\tau_\text{start})\} = \{0,\,0\}$
and/or
$\{w(\tau_\text{start}),\,\dot{w}(\tau_\text{start})\} = \{0,\,0\}$
and/or
$r_{M}(\tau_\text{start})= 0$.

\subsection{Dynamic cancellation of $\boldsymbol{\Lambda}$}
\label{subsec:Lambda-cancellation}

The calculational details of this section and the next
should not make us forget that the vector
fields of the model cancel the effective
cosmological constant $\Lambda$ exactly and without fine-tuning.

Indeed, the field equations \eqref{eq:reduced-field-eq-A-B}
and \eqref{eq:FRW-eqs-Friedmann-Einstein-matter} give nonzero
vector-field components, a Hubble parameter,
and a matter energy density
of the form \eqref{eq:asymp-sol} for coefficient $\gamma_{1}=0$,
\bsubeqs\label{eq:A0-B0-H-rM-sol}
\beqa\label{eq:A0-sol}
A_{0}(t) &=& \alpha_{0}\, (E_\text{Planck})^2\;t\,,\\[2mm]
\label{eq:B0-sol}
B_{0}(t) &=& \beta_{0}\, (E_\text{Planck})^2\;t\,,\\[2mm]
\label{eq:H-sol}
H(t) &=& \gamma_{0}\;t^{-1}\,,\\[2mm]
\label{eq:RM-sol}
\rho_{M}(t)
&=& \delta_{0}\;(E_\text{Planck})^2\;t^{-2}\,,
\eeqa
\esubeqs
where the overall normalization of $A_{0}$ and $B_{0}$ is irrelevant,
as only the ratio of the vector-field components enters
the action \eqref{eq:model-action}
according to Eqs.~\eqref{eq:def-FA-Q3A-FB-Q3B},
\eqref{eq:epsilon-Ansatz}, and \eqref{eq:Ansatz-A-B}.
The coefficients $\alpha_{0}$, $\beta_{0}$, $\gamma_{0}$, and $\delta_{0}$
in \eqref{eq:A0-B0-H-rM-sol} are not put in by hand but
appear dynamically. Specifically, the following
values have been calculated in Sec.~\ref{subsec:Expansion-coefficients}:
\bsubeqs\label{eq:values-sol}
\beqa
(\beta_{0}/\alpha_{0})^2 &=&
\frac{1}{2}\,\Lambda/(E_\text{Planck})^4
+ \sqrt{1+
\frac{1}{4}\,\Lambda^2/(E_\text{Planck})^8}
\,,
\\[2mm]
\gamma_{0} &=& 1/2\,,
\\[2mm]
\delta_{0} &=& 3/4\,.
\eeqa
\esubeqs

These particular fields give an exact cancellation
of $\Lambda$ appearing on the right-hand side of the generalized
FRW Eqs.~\eqref{eq:FRW-eqs-Friedmann-Einstein-matter},
\beqa\label{eq:Lambda-nullification}
\Lambda+\epsilon\Big( F_{A},\, F_{B}\Big)\,\Big|_\text{equil}
&=&
\Lambda+\epsilon\Big( (Q_{3A})^2,\, (Q_{3B})^2\Big)
\nonumber\\[2mm]
&=&
\Lambda+(E_\text{Planck})^4\;
\Big[(\alpha_{0}/\beta_{0})^2
-(\beta_{0}/\alpha_{0})^2\Big] =0\,,
\eeqa
where the definitions \eqref{eq:def-FA-Q3A-FB-Q3B},
\eqref{eq:epsilon-Ansatz}, and
\eqref{eq:FRW-Q3A-Q3B} have been used
for $F_{A,B}$, $\epsilon$, and $Q_{3A,B}$, respectively.
With the nullification \eqref{eq:Lambda-nullification},
the FRW Eqs.~\eqref{eq:FRW-eqs-Friedmann-Einstein-matter}
are solved to order $t^0$.

Including the higher-order terms of the asymptotic solution \eqref{eq:asymp-sol},  
also called the perfect-equilibrium solution later on, we can evaluate
the effective vacuum energy density of
what may be called the microscopic dark-energy component,
that is, the energy density not from standard matter
but from the initial (`bare') cosmological constant,
the vector fields, and the modified gravity.
A convenient definition for a spatially flat RW universe is as follows:
\beq\label{eq:def-rhoV}
\rho_{V\text{-micro}}(t) \equiv 3\,(E_\text{Planck})^2\,H(t)^2 -\rho_{M}(t)\,,
\eeq
which was simply denoted $\rho_V$
in Ref.~\cite{EmelyanovKlinkhamer2011-CCP1-FRW}.
The result from the asymptotic solution \eqref{eq:asymp-sol} is
\bsubeqs\label{eq:rV-leading-lim-rV-over-rM}
\beqa\label{eq:rV-leading}
\rho_{V\text{-micro}}(t)\,\Big|_\text{asymp.\;sol.} &=& \text{O}(t^{-5})\,,
\eeqa
which implies
\beqa\label{eq:lim-rV-over-rM}
\lim_{t\to\infty}\, \rho_{V\text{-micro}}(t)/\rho_{M}(t)\,\Big|_\text{asymp.\;sol.} &=& 0\,.
\eeqa
\esubeqs
Result \eqref{eq:lim-rV-over-rM} traces back
to the special properties of the $\epsilon$--function
\eqref{eq:epsilon-Ansatz} and was absent for the simpler models
of Refs.~\cite{EmelyanovKlinkhamer2011-CCP1-NEWTON,
EmelyanovKlinkhamer2011-CCP1-FRW}, which exhibited the behavior
$\rho_{V\text{-micro}}(t) \propto t^{-2}$.
Assuming the relevance of our model function \eqref{eq:epsilon-Ansatz}
to physics, the implication is that a new
mechanism is needed to explain the observed finite remnant
vacuum energy density of order $(\text{meV})^4$.

Expanding on the last remarks of the previous paragraph,
it is not difficult to see what the implications are for the present
energy-density ratio of dark energy and matter.
For the sake of the argument,
use $|\widetilde{\rho}_{V\text{-micro}}(t)| = t^{-4}$, which
may still be an overestimate as quantum-dissipative effects
can be expected to produce an exponential decrease
(cf. Ref.~\cite{KV2011-review} and paper [15] quoted therein).
A present cold-dark-matter
energy density of the order of the critical energy density
gives $\rho_\text{CDM}(t_{0}) \sim \;t_\text{Planck}^{-2}\;t_{0}^{-2}$,
for $t_\text{Planck}\equiv 1/E_\text{Planck}
\sim 10^{-42}\;\text{s}$
and $t_{0} \sim (c\,H_{0})^{-1} \sim 10^{17}\;\text{s}$.
The present energy-density ratio would then  be completely negligible,
$|\widetilde{\rho}_{V\text{-micro}}(t_{0})|/\rho_\text{CDM}(t_{0})
 \sim (t_\text{Planck}/t_{0})^2 \sim 10^{-118}$.
In fact, the ratio would already be extremely small near
the electroweak crossover:
$|\widetilde{\rho}_{V\text{-micro}}(t_\text{ew})|/\rho_\text{M}(t_\text{ew})
\sim (E_\text{ew}/E_\text{Planck})^4 \sim 10^{-60}$,
for $E_\text{ew} \sim \text{TeV}$ and
$t_\text{ew}\sim E_\text{Planck}/(E_\text{ew})^2$, as derived
from the spatially-flat Friedmann equation
with $\rho_\text{M}\sim (E_\text{ew})^4$.
With negligible $\widetilde{\rho}_{V\text{-micro}}(t)$ from the microscopic
variables ($A_\alpha$, $B_\alpha$, and effectively $\Lambda$),
further contributions to the vacuum energy density
$\rho_{V\text{-macro}}(t)$ may come
from phase transitions and mass effects of the macroscopic
standard-model fields.
As discussed in Ref.~\cite{KV2011-review}, the resulting
$\rho_{V\text{-macro}}(t)$ may decrease stepwise, approximately
as $t_\text{Planck}^{-2}\;t^{-2}$.

In conclusion, the exact solution \eqref{eq:A0-B0-H-rM-sol}
is of paramount importance,
especially if it is an attractor-type solution.
This attractor-type behavior will be discussed in the next section.

\section{Attractor-type solutions}
\label{sec:Attractor-type-solutions}

The present section is a direct follow-up of the previous one
and is also rather technical. In order to get an idea of the
attractor-type behavior, it is possible, in a first reading,
to consider only Sec.~\ref{subsec:Particular-class-exact-solutions}.

\subsection{Mathematical considerations}
\label{subsec:Mathematical-considerations}

The model of interest has an action-density
term $\epsilon(F_{A},F_{B})$ as given by \eqref{eq:epsilon-Ansatz}.
For completeness, two simpler models are discussed in the appendices:
in App.~A, the original Dolgov model~\cite{Dolgov1985-1997}
with just a $(Q_1)^2$ term in the
action density and, in App.~B, our previous
model~\cite{EmelyanovKlinkhamer2011-CCP1-FRW}
with a single $F_A$ term as defined in \eqref{eq:def-FA-Q3A}.

It turns out, however, that the first-order system of
differential equations for the $\epsilon(F_{A},F_{B})$ model
does not have the relatively simple structure as found in the
appendices, specifically,
Eqs.~\eqref{eq:linearized-system-Dolgov-model-vector-eq}
and \eqref{eq:linearized-system-ccp1-frw-model-vector-eq}.
Physically, the extra complications may be due to the
fact the $\epsilon(F_{A},F_{B})$ model is really
an $f(R)$ modified-gravity theory, which entails
higher-derivative field equations [in our case,
\eqref{eq:FRW-eqs-Einstein} has third-order derivatives of
$A_{0}(t)$, $B_{0}(t)$, and $H(t)$].

One possible way forward would be to rewrite this particular
modified-gravity theory as a scalar-tensor theory
(more precisely, a scalar-vector-tensor theory).
Instead, we prefer to adopt a low-tech (read brute-force)
approach by pushing the explicit solutions as far as possible.
This approach suffices to show the attractor-type behavior,
even though it lacks mathematical rigor compared
to the approach in the appendices. In fact, what would be
needed here is
the mathematical proof that the infinite sums in the expressions
of Sec.~\ref{subsec:Series-attractor-type-behavior} converge,
but we will simply
assume this to be the case, as has been done in most of the
literature on the subject (cf. Ref.~\cite[b]{Dolgov1985-1997}).
Still, awaiting this rigorous proof and the precise knowledge of
the attractor domain, we will only speak about
`attractor-type solutions' of the $\epsilon(F_{A},F_{B})$
model rather than `the attractor solution' \textit{tout court}.

\subsection{ODEs}
\label{subsec:ODEs}

The complete system of differential equations
from Sec.~\ref{subsec:Dimensionless ODEs}
can be written as follows:%
\bsubeqs\label{eq:ODEs-v-w-h-sh-rM-rewritten}
\beqa\label{eq:ODEs-v-rewritten}
0&=&
\ddot{v} + 3\,h\dot{v} -  6\,h^2\,v
+\big(\dot{v} +  3\,h\,v\big)\,
\frac{d}{d\tau}\ln\bigg|\frac{\partial \epsilondimless}
{\partial f_A}\bigg|\,,
\\[2mm]\label{eq:ODEs-w-rewritten}
0&=&
\ddot{w} + 3\,h\dot{w} -  6\,h^2\,w
+\big(\dot{w} +  3\,h\,w\big)\,
\frac{d}{d\tau}\ln\bigg|\frac{\partial\epsilondimless}
{\partial f_B}\bigg|\,,
\\[2mm]\label{eq:ODEs-h-rewritten}
0 &=&3\,h^2- \lambda - \epsilondimless - r_{M} -
3\,\big(\dot{h} + 3\,h^2\big)\,g_{AB} - 3\,h\,\dot{g}_{AB}\,,
\\[2mm]\label{eq:ODEs-sh-rewritten}
0 &=&
2\,\dot{h} + 3\,h^2 - \lambda - \epsilondimless +
\frac{1}{3}\,r_{M} +
\big(\dot{h} + 3\,h^2\big)\,g_{AB} - 2\,h\,\dot{g}_{AB} -\ddot{g}_{AB}\,,
\\[2mm]\label{eq:ODEs-rM-rewritten}
0 &=& \dot{r}_{M} + 4\,h\,r_{M}\,,
\eeqa
\esubeqs
where $\epsilondimless$, $f_{A}$, $f_{B}$, and $g_{AB}$
have already been defined in \eqref{eq:def-epsilondimless,fA,fB,gAB}.
We will now give several explicit analytic solutions of these ODEs.

\subsection{Particular class of exact solutions}
\label{subsec:Particular-class-exact-solutions}

The differential system \eqref{eq:ODEs-v-w-h-sh-rM-rewritten}
has the following class of exact solutions for $\tau>\tau_{0}\,$:
\bsubeqs\label{eq:exact-solution}
\beqa\label{eq:exact-solution-v}
v(\tau) &=& (\tau - \tau_{0})\,C_1
+ \frac{C_3}{C_2\,(\tau - \tau_{0})^{3/2}}\;,
\\[2mm]\label{eq:exact-solution-w}
w(\tau) &=& (\tau - \tau_{0})\,C_2
+C_{4}\; \frac{C_3}{C_1\,(\tau - \tau_{0})^{3/2}}\;,
\\[2mm]\label{eq:exact-solution-h}
h(\tau) &=& \frac{1/2}{(\tau - \tau_{0})}\;,
\\[2mm]\label{eq:exact-solution-rM}
r_M(\tau) &=& \frac{3/4}{(\tau - \tau_{0})^2}\;,
\eeqa
\esubeqs
with a real constant $ \tau_{0}\in \mathbb{R}$,
nonvanishing real constants $C_1,\,C_2\in \mathbb{R}\backslash \{0\}$,
a real constant $C_3\in \mathbb{R}$, and
a discrete constant $C_{4} \in \{-1,\,+1\}$.
The constants $\tau_{0}$, $C_3$, and $C_{4}$ are arbitrary.
The real ratio $C_1/C_2$ is
determined by the input cosmological constant $\lambda\,$ via a quartic
equation,%
\bsubeqs\label{eq:exact-solution-C1overC2-R12eq}
\beqa\label{eq:exact-solution-C1overC2}
C_1/C_2 &\equiv& R_{C}\,,\\[2mm]
\label{eq:exact-solution-R12eq}
\big(R_{C}\big)^4+\lambda\,\big(R_{C}\big)^2 &=& 1\,,
\eeqa
\esubeqs
as follows from, e.g., the generalized Friedmann equation
\eqref{eq:ODEs-h-rewritten}.
Hence, the number of free parameters
in \eqref{eq:exact-solution} is four:
$\tau_{0}$,  $(C_1\, C_2)$, $C_3$, and $C_{4}$.
The physically relevant parameters are, however, only
the ratio $C_3/(C_1\, C_2)$ and the relative sign $C_{4}$.

Observe that all solutions in \eqref{eq:exact-solution} give
for the effective vacuum energy density of the microscopic
degrees of freedom an exactly vanishing result,
\beqa\label{eq:rDE-zero}
r_{V\text{-micro}}(\tau)\,\Big|_{\tau_{0},\,C_1,\,C_2,\,C_3,\,C_{4}}^{C_1/C_2 = R_{C}}
&=& 0\,,
\eeqa
with definition $r_{V\text{-micro}}(\tau)\equiv 3\,h(\tau)^2 -r_M(\tau)$
from \eqref{eq:def-rhoV} and $R_{C}$
the positive or negative real solution
of \eqref{eq:exact-solution-R12eq}.
Result \eqref{eq:rDE-zero} also holds for the special case $C_3=0$,
which corresponds to the perfect-equilibrium solution
\eqref{eq:A0-B0-H-rM-sol} with constants \eqref{eq:values-sol}
and an arbitrary time-shift.

For the case of $C_3\ne 0$ and $C_{4} =-1$, the rescaled solutions
$v(\tau)/C_1$ and $w(\tau)/C_2$ in \eqref{eq:exact-solution}
are different at finite values of $\tau$,
specifically, $v(\tau)/C_1 - w(\tau)/C_2 \propto (\tau - \tau_{0})^{-3/2}$.
Still, both of these functions
$v(\tau)/C_1$ and $w(\tau)/C_2$ approach the same asymptotic
solution, the one from above, the other from below. This is precisely
the attractor-type behavior discussed in
Sec.~\ref{subsec:Mathematical-considerations}
and the two appendices (see also Refs.~\cite[(c)]{KV2008-2010} and
\cite{EmelyanovKlinkhamer2011-CCP1-NEWTON,EmelyanovKlinkhamer2011-CCP1-FRW}
for related numerical results).

\subsection{Series and attractor-type behavior}
\label{subsec:Series-attractor-type-behavior}

A generalized \textit{Ansatz}
for a nontrivial solution of \eqref{eq:ODEs-v-w-h-sh-rM-rewritten}
at $\tau\geq\tau_1 > 0$ is as follows:%
\bsubeqs\label{eq:general-Ansatz-functions}
\beqa\label{eq:general-Ansatz-function-v}
\tau^{-1}\,v(\tau)&=&
\frac{\big[v_1/\tau_1\big] +(\tau -\tau_1)^2}{1 +(\tau -\tau_1)^2}
+\frac{(\tau-\tau_1)\,\big[\dot{v}_1/\tau_1-v_1/\tau_1^2\big]
+(\tau -\tau_1)^2}
{1 +(\tau -\tau_1)^3}
\nonumber\\[1mm]&&
+\sum_{n=1}^{\infty}\, a_n \left(\frac{(\tau-\tau_1)^2}{\tau^3}\right)^n\,,
\\[2mm]
\label{eq:general-Ansatz-function-w}
(\beta_{0}\,\tau)^{-1}\,w(\tau)&=&
\frac{\big[w_1/(\beta_{0}\,\tau_1)]+(\tau -\tau_1)^2}{1 +(\tau -\tau_1)^2}
\nonumber\\[1mm]&&
+\frac{(\tau-\tau_1)\,\big[\dot{w}_1/(\beta_{0}\,\tau_1)
       -w_1/(\beta_{0}\,\tau_1^2)\big]+(\tau -\tau_1)^2}
      {1 +(\tau -\tau_1)^3}
\nonumber\\[1mm]&&
+\sum_{n=1}^{\infty}\, b_n \left(\frac{(\tau-\tau_1)^2}{\tau^3}\right)^n\,,
\\[2mm]
\label{eq:general-Ansatz-function-h}
2\,\tau\ h(\tau)&=&
\frac{\big[2\,\tau_1\,h_1] +(\tau -\tau_1)^2}{1 +(\tau -\tau_1)^2}
+\frac{(\tau-\tau_1)\,\big[2\,\tau_1\,\dot{h}_1+2\,h_1\big]
+(\tau -\tau_1)^2}{1 +(\tau -\tau_1)^3}
\nonumber\\[1mm]&&
+\sum_{n=1}^{\infty}\, c_n \left(\frac{(\tau-\tau_1)^2}{\tau^3}\right)^n\,,
\\[2mm]
\label{eq:general-Ansatz-function-rM}
(4/3)\,\tau^2\ r_{M}(\tau)&=&
\frac{\big[(4/3)\,\tau_1^2\,r_{M1}\big] +(\tau -\tau_1)^2}{1 +(\tau -\tau_1)^2}
\nonumber\\[1mm]&&
+\frac{(\tau-\tau_1)\,
\big[(8/3)\,\big(1-2\,\tau_1h_1\big)\,\tau_1\,r_{M1}\big]
+(\tau -\tau_1)^2}{1 +(\tau -\tau_1)^3}
\nonumber\\[1mm]&&
+\sum_{n=1}^{\infty}\, d_n \left(\frac{(\tau-\tau_1)^2}{\tau^3}\right)^n\,,
\eeqa
\esubeqs
with $\beta_{0}$ given by \eqref{eq:LO-coeffs}.
The seven constant parameters $v_1$, $\dot{v}_1$, $w_1$, $\dot{w}_1$,
$h_1$, $\dot{h}_1$, and $r_{M1}$
in  \eqref{eq:general-Ansatz-functions} represent the initial
values of the functions
and their first derivatives at $\tau = \tau_1$:%
\bsubeqs\label{eq:general-Ansatz-bcs}
\beqa
v(\tau_1) &=& v_1\,,\quad \dot{v}(\tau_1) = \dot{v}_1\,,\\[2mm]
w(\tau_1) &=& w_1\,,\quad \dot{w}(\tau_1) = \dot{w}_1\,,\\[2mm]
h(\tau_1) &=& h_1\,,\quad \dot{h}(\tau_1) = \dot{h}_1\,,\\[2mm]
r_{M}(\tau_1) &=& r_{M1}\,,
\eeqa
\esubeqs
where $r_{M}(\tau)$ requires only a single boundary condition
value as its ODE is first-order, the other ODEs being second-order.
These initial values must be sufficiently close
to those of the perfect-equilibrium solution,
given by \eqref{eq:exact-solution}
with $C_{1}=1$ and $C_{3}=\tau_{0}=0$.

Inserting the expansions \eqref{eq:general-Ansatz-functions}
into \eqref{eq:ODEs-v-w-h-sh-rM-rewritten}
gives values for the coefficients $a_n$, $b_n$, $c_n$, and $d_n$
in terms of the initial conditions $v_1$, $\,\ldots\,$, $r_{M1}$.
The expressions for these coefficients are rather bulky
(even for $\tau\gg\tau_1$)
and, here, we only indicate the dependence on the
initial conditions for the first few coefficients,
\bsubeqs\label{eq:general-Ansatz-coeff}
\beqa
a_1 &=& a_1(\tau_1,\,h_1,\,\dot{h}_1)\,,\\[2mm]
b_1 &=& b_1(\tau_1,\,h_1,\,\dot{h}_1)\,,\\[2mm]
c_i &=& c_i(\tau_1,\,h_1,\,\dot{h}_1)\,,\hspace*{18mm}\text{for}\;i=1,\,\ldots,\,5\,,\\[2mm]
d_1 &=& d_1(\tau_1,\,h_1,\,\dot{h}_1)\,,\\[2mm]
a_j &=& a_j(\tau_1,\,v_1,\,\dot{v}_1,\,h_1,\,\dot{h}_1)\,,
\quad\;\text{for}\;j=2,\,\ldots,\,7\,,\\[2mm]
b_j &=& b_j(\tau_1,\,w_1,\,\dot{w}_1,\,h_1,\,\dot{h}_1)\,,
\quad\text{for}\;j=2,\,\ldots,\,7\,,\\[2mm]
d_k &=& d_k(\tau_1,\,h_1,\,\dot{h}_1\,,r_{M1})\,,
\quad\;\;\;\;\text{for}\;k=2,\,\ldots,\,5\,.
\eeqa
\esubeqs
In the limit of large cosmic times (that is, large on the scale of the
Planck time, $\tau\gg\tau_1$), the corresponding
solution takes the following form:
\bsubeqs
\beqa
v(\tau) &=& \tau - 1 - c_1 + \text{O}\big(\tau^{-5}\big)\,,
\\[2mm]
w(\tau) &=& \beta_{0}\,\big(\tau-1-c_1\big)+\text{O}\big(\tau^{-5}\big)\,,
\\[2mm]
h(\tau) &=& \frac{1}{2\,\tau}
\Bigg[\sum\limits_{n =0}^4\,\frac{(1 + c_1)^n}{\tau^n}
+ \text{O}\Big(\frac{1}{\tau^5}\Big)\Bigg]\,,
\\[2mm]
r_M(\tau) &=& \frac{3}{4\,\tau^2}\Bigg[\sum\limits_{n = 0}^4\,(n + 1)\,
\frac{(1 + c_1)^n}{\tau^n} + \text{O}\Big(\frac{1}{\tau^5}\Big)\Bigg]\,.
\eeqa
\esubeqs
Extrapolating this result, we obtain the asymptotic
(perfect-equilibrium) solution,
\bsubeqs\label{eq:perfect-equilibrium-solution}
\beqa
v_\text{asymp}(\tau) &=& \tau - \widehat{\tau}_{1}\,,
\\[2mm]
w_\text{asymp}(\tau) &=& \beta_{0}\,\big(\tau - \widehat{\tau}_{1}\big)\,,
\\[2mm]
h_\text{asymp}(\tau) &=& \frac{1}{2}\,\big(\tau - \widehat{\tau}_{1}\big)^{-1}\,,
\\[2mm]
r_{M\text{asymp}}(\tau) &=& \frac{3}{4}\,
\big(\tau - \widehat{\tau}_{1}\big)^{-2}\,,
\eeqa
\esubeqs
where $\widehat{\tau}_{1} \equiv 1 + c_1$. Observe that,
apart from the overall time-shift $\widehat{\tau}_{1}$,
the obtained asymptotic solution is independent of the
initial conditions \eqref{eq:general-Ansatz-bcs}
encoded in the \textit{Ansatz} \eqref{eq:general-Ansatz-functions}.

The tentative conclusion is that different initial conditions give
different solutions, which, however, approach the
same asymptotic solution \eqref{eq:perfect-equilibrium-solution}.
Hence, there is an attractor-type behavior.
But, as explained in Sec.~\ref{subsec:Mathematical-considerations},
this conclusion needs to be proven rigorously
and the proper attractor domain needs to be determined.

\section{Second-order perturbations}
\label{sec:Second-order-perturbations}

We, now, turn to localized perturbations of the metric tensor
field and the two vector fields. Denoting the four spacetime
coordinates $(x_1,\,x_2,\,x_3,\,t)$ collectively as $x$, we consider
the tensor field
\beq
g_{\alpha\beta}(x) = g_{\alpha\beta}(t) +\widehat{h}_{\alpha\beta}(x),
\eeq
with  the metric $g_{\alpha\beta}(t)$ of the flat
RW spacetime \eqref{eq:RW-metric}
and $\widehat{h}_{\alpha\beta}(x)$ the perturbation
\mbox{($|\widehat{h}_{\alpha\beta}| \ll 1$).}
On small scales, the relevant background metric is the
standard Minkowski metric
$\eta_{\alpha\beta}=\textrm{diag}(1,\,-1,\,-1,\,-1)$.
In addition, we consider the two vector fields
\bsubeqs
\beqa
A_{\alpha}(x) &=& A_{\alpha}(t) + \delta{A}_{\alpha}(x)\,,\\[2mm]
B_{\alpha}(x) &=& B_{\alpha}(t) + \delta{B}_{\alpha}(x)\,,
\eeqa
\esubeqs
with
$A_{\alpha}(t) = A_{0}(t)\,\delta_{\alpha}^0$,
$B_{\alpha}(t) = B_{0}(t)\,\delta_{\alpha}^0$,
$|\delta{A}_{\alpha}| \ll |A_{0}|$, and
$|\delta{B}_{\alpha}| \ll |B_{0}|$.

This section is highly technical and, in a first reading,
it is possible to skip ahead to
Sec.~\ref{Standard-local-Newtonian-dynamics} with the
main physics result of this section.

\subsection{Variation of the vector-field Lagrange density}
\label{subsec:lagrange-density}

The Lagrange density of the vector fields is given by
$\mathcal{L}_{A,B} = \Lambda+ \epsilon(F_{A},F_{B})$,
where the effective cosmological constant $\Lambda$
has been included for convenience.
To second order, the perturbed Lagrange density reads
\beqa\label{eq:Lagrange-AB-012}
\mathcal{L}_{A,B}^\text{(perturb.)} &=& \mathcal{L}_{A,B}^{(0)} +
\mathcal{L}_{A,B}^{(1)} + \mathcal{L}_{A,B}^{(2)}\,,
\eeqa
with
\bsubeqs
\beqa
\mathcal{L}_{A,B}^{(0)} &=& \Lambda+\epsilon(F_{A},F_{B})\,,\\[2mm]
\mathcal{L}_{A,B}^{(1)} &=&
\frac{\partial\,\epsilon}{\partial\,F_{A}}\,\delta^{(1)}F_{A} +
\frac{\partial\,\epsilon}{\partial\,F_{B}}\,\delta^{(1)}F_{B}\,,
\\[2mm]
\mathcal{L}_{A,B}^{(2)} &=&
\frac{\partial\,\epsilon}{\partial\,F_{A}}\,\delta^{(2)}F_{A}
+ \frac{\partial\,\epsilon}{\partial\,F_{B}}\,\delta^{(2)}F_{B}
+ \frac{1}{2}\,\frac{\partial^{\,2}\,\epsilon}{\partial\,F_{A}^2}\,
\Big(\delta^{(1)}F_{A}\Big)^2
\nonumber\\[1mm]&&
+ \frac{\partial^{\,2}\,\epsilon}{\partial\,F_{A}\partial\,F_{B}}\,
\delta^{(1)}F_{A}\,\delta^{(1)}F_{B}
+ \frac{1}{2}\,\frac{\partial^{\,2}\,\epsilon}{\partial\,F_{B}^2}\,
\Big(\delta^{(1)}F_{B}\Big)^2\,.
\eeqa
\esubeqs
The first- and second-order variations of $Q_A$ and $F_{A}$ are
\bsubeqs\label{eq:perturbations-of-QA-and-FA}
\beqa
\delta^{(1)}Q_{3A} &=& \delta{A}_{\alpha}^{;\alpha} +
\widehat{h}^{\alpha\beta}\,A_{\alpha;\beta} - g^{\alpha\beta}\delta^{(1)}
\Gamma_{\alpha\beta}^{\gamma}A_{\gamma}\,,
\\[2mm]
\delta^{(2)}Q_{3A} &=& \widehat{h}^{\alpha\beta}\,\delta{A}_{\alpha;\beta} -
g^{\alpha\beta}\delta^{(1)}
\Gamma_{\alpha\beta}^{\gamma}\delta{A}_{\gamma}
- \widehat{h}^{\alpha\beta}\,\delta^{(1)}\Gamma_{\alpha\beta}^{\gamma}A_{\gamma}
- g^{\alpha\beta}\,\delta^{(2)}\Gamma_{\alpha\beta}^{\gamma}A_{\gamma}\,,
\eeqa
\beqa
\delta^{(1)}F_{A} &=& 2Q_{3A}\delta^{(1)}Q_{3A} - \frac{1}{2}R
\Big(2A^{\alpha}\delta{A}_{\alpha} +
\widehat{h}^{\alpha\beta}\,A_{\alpha}A_{\beta}\Big)
- \frac{1}{2}A^2\delta^{(1)}R\,,
\\[2mm]
\delta^{(2)}F_{A} &=& 2Q_{3A}\delta^{(2)}Q_{3A} +
(\delta^{(1)}Q_{3A})^2
- \frac{1}{2}R\Big(\delta{A}^{\alpha}\delta{A}_{\alpha} +
2\widehat{h}^{\alpha\beta}\,A_{\alpha}\delta{A}_{\beta}\Big)
\nonumber\\[2mm]&&
- \frac{1}{2}\delta^{(1)}R\Big(2A^{\alpha}\delta{A}_{\alpha} +
\widehat{h}^{\alpha\beta}\,A_{\alpha}A_{\beta}\Big) -
\frac{1}{2}A^2\delta^{(2)}R\,.
\eeqa
\esubeqs
Replacing $A_{\alpha}$ and $\delta{A}_{\alpha}$
in ($\ref{eq:perturbations-of-QA-and-FA}$) by
$B_{\alpha}$ and $\delta{B}_{\alpha}$
gives the first- and second-order variations of $Q_B$ and $F_{B}$.

For future use, we rewrite $\mathcal{L}_{A,B}^{(1)}$
and $\mathcal{L}_{A,B}^{(2)}$ in dimensionless form,
\bsubeqs
\beqa
\mathcal{L}_{A,B}^{(1)} &=&
  Q_{3A0}^2\,\frac{\partial\,\epsilon}{\partial\,F_{A}}\delta^{(1)}f_A
+ Q_{3B0}^2\,\frac{\partial\,\epsilon}{\partial\,F_{B}}\delta^{(1)}f_B\,,
\\[2mm]
\mathcal{L}_{A,B}^{(2)} &=&
  Q_{3A0}^2\,\frac{\partial\,\epsilon}{\partial\,F_{A}}\delta^{(2)}f_A
+ Q_{3B0}^2\,\frac{\partial\,\epsilon}{\partial\,F_{B}}\delta^{(2)}f_B
+ \frac{1}{2}\,Q_{3A0}^4\,\frac{\partial^{\,2}\,\epsilon}{\partial\,F_{A}^2}
\,\Big(\delta^{(1)}f_A\Big)^2
\nonumber\\[1mm]&&
+ Q_{3A0}^2Q_{3B0}^2\,\frac{\partial^{\,2}\epsilon}
{\partial\,F_{A}\partial\,F_{B}}\delta^{(1)}f_A\delta^{(1)}f_B
+ \frac{1}{2}\,Q_{3B0}^4\,\frac{\partial^{\,2}\,\epsilon}{\partial\,F_{B}^2}
\,\Big(\delta^{(1)}f_B\Big)^2\,,
\eeqa
\esubeqs
where $\delta^{(1)}f_{A,B}$ and $\delta^{(2)}f_{A,B}$ correspond to
$\delta^{(1)}F_{A,B}$ and $\delta^{(2)}F_{A,B}$ expressed in terms
of dimensionless variables $\overline{v}_{\alpha}$,
$\overline{w}_{\alpha}$ and
$\widehat{v}_{\alpha}$, $\widehat{w}_{\alpha}$.
These dimensionless variables are defined as follows:%
\bsubeqs\label{eq:def-v-w-vhat-what-xhat}
\beqa
\overline{v}_{\alpha}(t) &\equiv& \frac{1}{Q_{3A0}}\,A_{\alpha}(t)\,,\quad\;\,
\overline{w}_{\alpha}(t) \equiv \frac{1}{Q_{3B0}}\,B_{\alpha}(t)\,,
\\[2mm]
\widehat{v}_{\alpha}(x) &\equiv& \frac{1}{Q_{3A0}}\,\delta{A}_{\alpha}(x)
\,,\quad
\widehat{w}_{\alpha}(x)   \equiv \frac{1}{Q_{3B0}}\,\delta{B}_{\alpha}(x)\,,
\\[2mm]\label{eq:def-xhat}
\widehat{z}_{\alpha} &\equiv& \widehat{v}_{\alpha} - \widehat{w}_{\alpha}\,,
\eeqa
\esubeqs
with dimensional constants $Q_{3A0}$ and $Q_{3B0}$.
In \eqref{eq:def-xhat}, we have added the
definition of $\widehat{z}_{\alpha}$,
which will be used extensively in the next subsections.
Note also that, in the above definitions,
the background fields are distinguished by a bar
and the perturbation fields by a hat.

\subsection{Equations for the vector-field perturbations}
\label{subsec:equations-of-vector-field-perturbations}

The equations of the vector-field perturbations are
\bsubeqs
\beqa\label{eq:vector-field-eq-for-perturbation-A}
\hspace*{-6mm}
\partial_{\alpha}\bigg(\overline{q}_{3A}\,
\bigg[
 Q_{3A0}^2\,\frac{\partial^{\,2}\,\epsilon}{\partial\,F_{A}^2}\,\delta{f}_A
+Q_{3B0}^2\,\frac{\partial^{\,2}\,\epsilon}{\partial\,F_{A}\partial\,F_{B}}\,
\delta{f}_B\bigg]
+ \frac{\partial\,\epsilon}{\partial\,F_{A}}\delta{q}_{3A}\bigg)
\nonumber\\[2mm]
\hspace*{-6mm}
+ \frac{1}{2}\,\frac{\partial\,\epsilon}{\partial\,F_{A}}
\Big(\delta{R}\,\overline{v}_{\alpha}+ R\,\widehat{v}_{\alpha}\Big)
+ \frac{1}{2}\,R\,
\bigg(
 Q_{3A0}^2\,\frac{\partial^{\,2}\,\epsilon}{\partial\,F_{A}^2}\,\delta{f}_A
+Q_{3B0}^2\,\frac{\partial^{\,2}\,\epsilon}{\partial\,F_{A}\partial\,F_{B}}\,
\delta{f}_B
\bigg)\,\overline{v}_{\alpha} &=& 0,
\eeqa
\beqa
\label{eq:vector-field-eq-for-perturbation-B}
\hspace*{-6mm}
\partial_{\alpha}\bigg(\overline{q}_{3B}\,
\bigg[
 Q_{3B0}^2\,\frac{\partial^{\,2}\,\epsilon}{\partial\,F_{B}^2}\,\delta{f}_B
+Q_{3A0}^2\,\frac{\partial^{\,2}\,\epsilon}{\partial\,F_{A}\partial\,F_{B}}\,
\delta{f}_A\bigg]
+ \frac{\partial\,\epsilon}{\partial\,F_{B}}\,\delta{q}_{3B}\bigg)
\nonumber\\[2mm]
\hspace*{-6mm}
+ \frac{1}{2}\frac{\partial\,\epsilon}{\partial\,F_{B}}
\Big(\delta{R}\,\overline{w}_{\alpha} + R\,\widehat{w}_{\alpha}\Big)
+ \frac{1}{2}\,R\,\bigg(
Q_{3B0}^2\,\frac{\partial^{\,2}\,\epsilon}{\partial\,F_{B}^2}\,\delta{f}_B +
Q_{3A0}^2\,\frac{\partial^{\,2}\,\epsilon}{\partial\,F_{A}\partial\,F_{B}}\,
\delta{f}_A\bigg)\,\overline{w}_{\alpha} &=& 0\,,
\eeqa
\esubeqs
with $\delta{f}_{A,B} \equiv \delta^{(1)}f_{A,B}$,
$\delta{q}_{3A,3B} \equiv \delta^{(1)}\overline{q}_{3A,3B}$, and
$\delta{R} \equiv \delta^{(1)}R$. Furthermore, we have
$\overline{q}_{3A} = \dot{\overline{v}}_{0} + 3\,h\,\overline{v}_{0}$ and
$\overline{q}_{3B} = \dot{\overline{w}}_{0} + 3\,h\,\overline{w}_{0}$.
Note that the above equations for the perturbations $\widehat{v}_{\alpha}$
and $\widehat{w}_{\alpha}$ carry a third derivative of the metric
perturbation $\widehat{h}_{\alpha\beta}$, since $\delta{f}_A$ and $\delta{f}_B$
contain $\delta{R}$, which already has a second
derivative of $\widehat{h}_{\alpha\beta}$.

Using the background vector-field
Eqs.~($\ref{eq:vector-field-eq-A}$)
and ($\ref{eq:vector-field-eq-B}$),
the perturbation
Eqs.~($\ref{eq:vector-field-eq-for-perturbation-A}$)
and ($\ref{eq:vector-field-eq-for-perturbation-B}$)
can be reduced to
\bsubeqs
\beqa\label{eq:simplified-vector-field-equation-for-perturbation-A}
\partial_{\alpha}
\bigg(\delta{\Omega}_A +\frac{\delta{q}_{3A}}{\overline{q}_{3A}}\bigg)
+ \frac{1}{2\overline{q}_{3A}}
\bigg(R\,\widehat{v}_{\alpha}+ \bigg[\delta{R} -
R\,\frac{\delta{q}_{3A}}{\overline{q}_{3A}}\bigg]\,
\overline{v}_{\alpha}\bigg) &=& 0\,,
\\[2mm]\label{eq:simplified-vector-field-equation-for-perturbation-B}
\partial_{\alpha}
\bigg(\delta{\Omega}_B +\frac{\delta{q}_{3B}}{\overline{q}_{3B}}\bigg)
+ \frac{1}{2\overline{q}_{3B}}\bigg(R\,\widehat{w}_{\alpha}
+ \bigg[\delta{R} -
R\,\frac{\delta{q}_{3B}}{\overline{q}_{3B}}\bigg]\,
\overline{w}_{\alpha}\bigg) &=& 0\,,
\eeqa
\esubeqs
with definitions
\bsubeqs\label{eq:def-OmegaA-OmegaB}
\beqa\label{eq:def-OmegaA}
\delta{\Omega}_A &\equiv&
\bigg(\frac{\partial\,\epsilon}{\partial\,F_{A}}\bigg)^{-1}\;
\bigg(\frac{\partial^{\,2}\,\epsilon}{\partial\,F_{A}^2}\,\delta{F}_A +
\frac{\partial^{\,2}\,\epsilon}{\partial\,F_{A}\partial\,F_{B}}\,\delta{F}_B
\bigg)\,,
\\[2mm]
\label{eq:def-OmegaB}
\delta{\Omega}_B &\equiv&
\bigg(\frac{\partial\,\epsilon}{\partial\,F_{B}}\bigg)^{-1}\;
\bigg(\frac{\partial^{\,2}\,\epsilon}{\partial\,F_{B}^2}\,\delta{F}_B +
\frac{\partial^{\,2}\,\epsilon}{\partial\,F_{A}\partial\,F_{B}}\,\delta{F}_A
\bigg)\,.
\eeqa
\esubeqs

Taking $\overline{v}_{\alpha}(t) = \overline{w}_{\alpha}(t) =
\zeta_{\alpha}(t)$
for $\zeta_{\alpha}(t) = (\zeta(t),\,0,\,0,\,0)$, we find
$\overline{q}_{3A} = \overline{q}_{3B} \equiv \overline{q}_{3} = \dot{\zeta} + 3\,h\,\zeta$.
Subtracting
($\ref{eq:simplified-vector-field-equation-for-perturbation-B}$) from
($\ref{eq:simplified-vector-field-equation-for-perturbation-A}$)
then gives
\beqa\label{eq:hat-x}
\partial_{\alpha}\,\Big(\Xi\,
\Big[2\,\overline{q}_{3}\,\nabla^{\beta}\widehat{z}_{\beta}
-R\,\zeta^{\beta}\,\widehat{z}_{\beta}\Big]
+\frac{1}{\overline{q}_{3}}\,\nabla^{\beta}\,\widehat{z}_{\beta}\Big)
+\frac{1}{2\,\overline{q}_{3}}\,R\,\widehat{z}_{\alpha}
-\frac{1}{2\,(\overline{q}_{3})^2}\,R\,
\zeta_{\alpha}\,\nabla^{\beta}\widehat{z}_{\beta} &=& 0\,,
\eeqa
where $\widehat{z}_{\alpha}$ has already been defined in
\eqref{eq:def-v-w-vhat-what-xhat}
and
\beqa\label{eq:Xi}
\Xi &\equiv&
\frac{(Q_{3A})^2}{(\overline{q}_{3})^2}\;
\bigg[\bigg(
\frac{\partial\,\epsilon}{\partial\,F_{A}}\bigg)^{-1}\frac{\partial^{\,2}\epsilon}
{\partial\,F_{A}^2}
 - \bigg(\frac{\partial\,\epsilon}{\partial\,F_{B}}\bigg)^{-1}
 \frac{\partial^{\,2}\,\epsilon}{\partial\,F_{A}\partial\,F_{B}}\bigg]
\nonumber\\[2mm]
\phantom{\Xi} &\equiv&
\frac{(Q_{3B})^2}{(\overline{q}_{3})^2}\;
\bigg[\bigg(
\frac{\partial\,\epsilon}{\partial\,F_{B}}\bigg)^{-1}\frac{\partial^{\,2}
\epsilon}{\partial\,F_{B}^2}
 - \bigg(\frac{\partial\,\epsilon}{\partial\,F_{A}}\bigg)^{-1}
 \frac{\partial^{\,2}\,\epsilon}{\partial\,F_{A}\partial\,F_{B}}\bigg]\,.
\eeqa

Notice that $\widehat{z}_{\alpha}$
in \eqref{eq:hat-x} is not coupled to the
metric perturbation $\widehat{h}_{\alpha\beta}\,$:
$\widehat{z}_{\alpha}$ depends only on the functions $\zeta(t)$ and
$H(t)$ from the background fields,
together with the initial conditions for $\widehat{v}_{\alpha}$
and $\widehat{w}_{\alpha}$. This result follows from the symmetry
properties of the function $\epsilon(F_{A},F_{B})$.
Substituting the $\epsilon$ function \eqref{eq:epsilon-Ansatz}
into \eqref{eq:Theta}, we find
\beqa\label{eq:Theta-final}
\Xi &=& -\frac{1}{(\overline{q}_{3})^2 - (1/2)\,R\,\zeta^2}\;\,.
\eeqa

In the perfect-equilibrium state with Hubble parameter
$H(t)=1/2\:t^{-1}$ (implying $R=0$)
and constant values of $Q_{3A}$
and $Q_{3B}$ (as mentioned in the last sentence of
Sec.~\ref{subsec:Reduced-vector-field-equations}),
Eqs.~\eqref{eq:hat-x} and \eqref{eq:Theta-final}
give the following final equation for
$\widehat{z}_{\alpha}\equiv\widehat{v}_{\alpha}-\widehat{w}_{\alpha}\,$:%
\beqa\label{eq:hat-z-final}
\nabla_{\alpha}\,\nabla^{\beta}\,\widehat{z}_{\beta}\,
\Big|_\text{equil.\;background}
&=&
\partial_{\alpha}\big[t^{-3/2}\,\partial^{\beta}
\big(t^{3/2}\,\widehat{z}_{\beta}\big)\big]
=0\,.
\eeqa
For perturbation fields which are analytic and of finite support
($\widehat{v}_{\alpha}=\widehat{w}_{\alpha}=0$ for $t\in [0,\,T]$ and
$|\vec{x}| \geq R$), the solution is trivial and%
\beqa\label{eq:hat-z-final-zero}
\widehat{z}_{\alpha}\,
\Big|_\text{equil.\;background}^\text{local\;perturb.}
&=& 0\,.
\eeqa
In other words, the two linear vector-field perturbations
turn out to be equal,
$\delta{A}_{\alpha}(x)= \delta{B}_{\alpha}(x)$, which is
the same result as obtained in
Ref.~\cite{EmelyanovKlinkhamer2011-CCP1-NEWTON}
by different methods.
The explanation of \eqref{eq:hat-z-final-zero} is simple:
the localized perturbation fields
$\widehat{v}_{\alpha}$ and $\widehat{w}_{\alpha}$
obey the same equation and their boundary conditions
over an exterior region are also the same (zero, in fact).

\subsection{Energy-momentum tensor of the vector-field perturbations}
\label{subsec:energy-momentum-tensor-of-vector-field-perturbations}

The linear perturbation of the energy-momentum tensor
of the vector fields is given by the following expression
(only the arguments $\widehat{h}$,
$\widehat{v}$, and $\widehat{w}$ are shown explicitly
on the left-hand side):
\beqa\label{eq:Theta}
\hspace*{-12mm}&&
\Theta_{\alpha\beta}[\,\widehat{h},\,\widehat{v},\,\widehat{w}\,]
=
\big(\Lambda+\epsilon\big)\;\widehat{h}_{\alpha\beta}
\nonumber\\[2mm]
\hspace*{-12mm}&&
+\textstyle{\frac{1}{2}}\,(\mu_{A0}-\mu_{B0})\;
\Big(\big(2\,\overline{q}_{3}\,\nabla^{\lambda}\widehat{z}_{\lambda}
- R\,\zeta^{\lambda}\,\widehat{z}_{\lambda}\big)\,g_{\alpha\beta}
+ 2\,R_{\alpha\beta}\,\zeta^{\lambda}\,\widehat{z}_{\lambda}
- R\,\big(\zeta_{\alpha}\,\widehat{z}_{\beta}
  + \zeta_{\beta}\,\widehat{z}_{\alpha}\big)\Big)
\nonumber\\[2mm]
\hspace*{-12mm}&&
+\Big(\nabla_{\alpha}\nabla_{\beta}- g_{\alpha\beta}\,\nabla^2\Big)\,
 \Big(2\mu_{B0}\,\zeta^{\lambda}\,\widehat{z}_{\lambda}
+ \nu_{B0}\,\big(2\,\overline{q}_{3}\,\nabla^{\lambda}\widehat{z}_{\lambda}
-R\,\zeta^{\lambda}\,\widehat{z}_{\lambda}\big)\,\zeta^2\Big)
\nonumber\\[2mm]
\hspace*{-12mm}&&
+ \textstyle{\frac{1}{2}}\,(\nu_{A0}-\nu_{B0})\;
\Big(2\,\overline{q}_{3}\,\nabla^{\lambda}\,\widehat{z}_{\lambda}
-R\,\zeta^{\lambda}\,\widehat{z}_{\lambda}\Big)\,
\Big(R_{\alpha\beta}\,\zeta^2-R\,\zeta_{\alpha}\zeta_{\beta}\Big)\,,
\eeqa
with definitions
\bsubeqs
\beqa
\mu_{A} &\equiv& (Q_{3A})^2\,\frac{\partial\,\epsilon}{\partial\,F_{A}},\quad
\mu_{B} \equiv (Q_{3B})^2\,\frac{\partial\,\epsilon}{\partial\,F_{B}}\,,
\\[2mm]
\nu_{A} &\equiv& (Q_{3A})^2\,\Big((Q_{3A})^2\,\frac{\partial^{\,2}\epsilon}
{\partial\,F_{A}^2} + (Q_{3B})^2\,\frac{\partial^{\,2}\epsilon}
{\partial\,F_{A}\partial\,F_{B}}\Big)\,,
\\[2mm]
\nu_{B} &\equiv& (Q_{3B})^2\,\Big((Q_{3B})^2\,\frac{\partial^{\,2}\epsilon}
{\partial\,F_{B}^2} + (Q_{3A})^2\,\frac{\partial^{\,2}\epsilon}
{\partial\,F_{A}\partial\,F_{B}}\Big)\,,
\eeqa
\esubeqs
so that $\mu_{A0} + \mu_{B0} = 0$ and $\nu_{A0} + \nu_{B0} = 0$
for the special function \eqref{eq:epsilon-Ansatz}
and the perfect-equilibrium background fields
with a Ricci-flat spacetime ($R=0$).

Manifestly,
$\Theta_{\alpha\beta}[\,\widehat{h},\,\widehat{v},\,\widehat{w}\,]$
does not contain derivatives of the metric perturbation
$\widehat{h}_{\alpha\beta}$
and depends only on the difference between the vector-field perturbations,
$\widehat{z}_{\alpha}$ as defined by \eqref{eq:def-xhat}.
These results rely on the symmetry properties
($\ref{eq:property-1}$) and ($\ref{eq:property-2}$)
of the special $\epsilon(F_{A},F_{B})$ function \eqref{eq:epsilon-Ansatz}
and on the fact that $F_{A}$ and $F_{B}$
are quadratic with respect to the vector fields and that the background
fields evolve identically as mentioned below \eqref{eq:def-OmegaB}.

\subsection{Standard local Newtonian dynamics}
\label{Standard-local-Newtonian-dynamics}
\vspace*{-2mm}

With the results of the previous two subsections,
we can, at last, turn to the physical question of interest: the
gravitational self-interaction of small (noncosmological) systems.
This has been discussed extensively in our
previous article~\cite{EmelyanovKlinkhamer2011-CCP1-NEWTON},
so we can be brief.

The linear equation for the weak gravitational field from a localized
matter distribution is then%
\bsubeqs\label{eq:weak-grav-eq-h-S-T}
\beqa\label{eq:weak-grav-eq-h}
&&\square\,\widehat{h}_{\alpha\beta} +16\,\pi\,G\,S_{\alpha\beta}=0\,,
\\[1mm]\label{eq:weak-grav-eq-S}
&&S_{\alpha\beta}\equiv
T_{\alpha\beta} -\frac{1}{2}\,\eta_{\alpha\beta}\,
\eta^{\gamma\delta}\,T_{\gamma\delta}\,,
\\[1mm]\label{eq:weak-grav-eq-T}
&&T_{\alpha\beta}=T_{\alpha\beta}^\text{\,(matter)}+\Theta_{\alpha\beta}\,,
\eeqa\esubeqs
where the harmonic gauge,
$\partial_{\alpha} \widehat{h}^{\alpha}_{\;\;\beta}
= (1/2)\,\partial_{\beta}\,\widehat{h}^{\alpha}_{\;\;\alpha}\,$,
has been used to simplify
the standard derivative term on the left-hand side
of \eqref{eq:weak-grav-eq-h}, with d'Alembertian
$\square\equiv \eta^{\alpha\beta}\,\partial_{\alpha}\partial_{\beta}$.
The only new contribution appears as the second term
on the right-hand side of \eqref{eq:weak-grav-eq-T}
and has been given in \eqref{eq:Theta}.

Several comments are in order. First, note that
the background fields $\overline{v}_{\alpha}(\tau)$
and $\overline{w}_{\alpha}(\tau)$
are such that the $\Lambda+\epsilon$
term in \eqref{eq:Theta} vanishes for the perfect-equilibrium background;
see, in particular, the derivation \eqref{eq:Lambda-nullification}.
Second, recall that the energy-momentum tensor $\Theta_{\alpha\beta}$
of the perturbations depends only on
the metric perturbation $\widehat{h}_{\alpha\beta}$
(but not its derivatives)
and the difference of the vector-field perturbations.
Specifically, the behavior is as follows,
in a symbolic notation: $\Theta_{\alpha\beta}
\big[\,\widehat{h},\,\widehat{v},\,\widehat{w}\big] =
\Theta_{\alpha\beta}\big[\,\widehat{h},\,(\widehat{v} - \widehat{w}),\,
(\nabla+\nabla^2+\nabla^3)\,(\widehat{v} - \widehat{w})\big]$.
The main input for this result is that the normalized
background vector fields evolve identically,
$\overline{v}_{\alpha} = \overline{w}_{\alpha} = \zeta_{\alpha}(t)$
for $t\to\infty$.
But this is precisely what was found in
Sec.~\ref{sec:Asymptotic-solution}.
The evolution of $\widehat{z}_{\alpha}
\equiv \widehat{v}_{\alpha} - \widehat{w}_{\alpha}$
is, therefore, not affected by the metric perturbation $\widehat{h}_{\alpha\beta}$
(at least, to the linear order in perturbation theory considered).
Moreover, \eqref{eq:hat-z-final-zero} states that $\widehat{z}_{\alpha}$
vanishes due to the boundary conditions at infinity
(the energy density of the matter perturbation being localized
in space and time).

With $\Lambda+\epsilon=0$ and $\widehat{z}_{\alpha}=0$
nullifying \eqref{eq:Theta}, the conclusion is that the nonstandard
term in \eqref{eq:weak-grav-eq-h-S-T} drops out,
\bsubeqs\label{eq:ThetaIsZero-standardGR}
\beqa\label{eq:ThetaIsZero}
\Big[\Theta_{\alpha\beta}[\,\widehat{h},\,\widehat{v},\,\widehat{w}\,]\,
\Big]_\text{equil.\;background}^\text{local\;perturb.}=0\,,
\eeqa
and that the linear weak-gravity field equation (in harmonic gauge)
equals the one of general relativity,
\beqa\label{eq:standardGR}
\Big[\square\,\widehat{h}_{\alpha\beta}
+16\,\pi\,G\,S_{\alpha\beta}^\text{\,(matter)}\,
\Big]_\text{equil.\;background}^\text{local\;perturb.}=0\,,
\eeqa
with the standard-matter source term
\beqa
&&S_{\alpha\beta}^\text{\,(matter)}\equiv
T_{\alpha\beta}^\text{\,(matter)}
-\frac{1}{2}\,\eta_{\alpha\beta}\,
\eta^{\gamma\delta}\,T_{\gamma\delta}^\text{\,(matter)}\,.
\eeqa
\esubeqs
As mentioned before, these results hold for perfect-equilibrium
background fields
[given by \eqref{eq:RW-metric}, \eqref{eq:Ansatz-A-B},
and \eqref{eq:A0-B0-H-rM-sol} in dimensional form
or \eqref{eq:perfect-equilibrium-solution} in dimensionless form],
which have dynamically canceled the cosmological constant $\Lambda$
(see Sec.~\ref{subsec:Lambda-cancellation}).
Recall that the main cosmological constant problem, CCP1
as formulated in Sec.~\ref{sec:intro}, is precisely
concerned with the dynamic cancellation of $\Lambda$
in the \emph{equilibrium} state of the quantum vacuum.
The study of small self-gravitating systems in a nonequilibrium
background (even if this background rapidly approaches the
equilibrium state, as discussed in the penultimate paragraph of
Sec.~\ref{subsec:Lambda-cancellation}),
lies outside the scope of the present article~\cite{Endnote-graviton-mass}.

Equation \eqref{eq:standardGR} shows, in particular, that
the standard Newtonian law of gravity (i.e., the Poisson equation)
holds for local nonrelativistic
matter distributions such as the Solar System or the Galaxy.
This implies that the constant $G$ in \eqref{eq:standardGR}, which
traces back to the original action \eqref{eq:model-action-EPlanck},
can be identified with Newton's gravitational coupling constant,
\beq
G=G_{N} = 6.6743(7) \times
10^{-11}\;\text{m}^{3}\;\text{kg}^{-1}\;\text{s}^{-2}\,,
\eeq
where the numerical value has been taken from the CODATA--2006
compilation~\cite{MohrTaylorNewell2008}.
The corresponding numerical value of the gravitational
energy scale defined in \eqref{eq:EPlanck} is then the usual one,
$E_\text{Planck} \approx 2.44\times 10^{18}\:\text{GeV}$.

The present article,
just as its predecessor~\cite{EmelyanovKlinkhamer2011-CCP1-NEWTON},
only considers the linear theory of small self-gravitating systems.
It remains to be seen whether or not the present setup
reproduces locally the standard nonlinear theory, i.e., general relativity.

\section*{\hspace*{-4.5mm}ACKNOWLEDGMENTS}
\vspace*{-0mm}\noindent
It is a pleasure to thank the referee for helpful remarks.

\begin{appendix}
\section{Attractor solution in a model with a $\boldsymbol{(Q_1)^2}$ term}
\label{app:attractor-mathematics-Dolgov-model}

\subsection{ODEs and new variables}
\label{app-subsec:ODEs-and-new-variables-ccp1-Dolgov-model}

The original Dolgov model~\cite{Dolgov1985-1997},
with a single massless vector field $A_{\alpha}(x)$ and
a positive cosmological constant $\Lambda$,
is defined by the action \eqref{eq:model-action},
setting $B_{\alpha}(x) \equiv 0$ and having
a vacuum-energy-density term $\epsilon_{D}$ based on a different
contraction of the vector-field derivatives,
\bsubeqs\label{eq:DM-epsilon-Q1}
\beqa\label{eq:DM-epsilon}
\epsilon_{D}&=&(Q_1)^2\,,\\[2mm]
\label{eq:DM-Q1}
(Q_1)^2 &\equiv& A_{\alpha;\beta}\,A^{\alpha;\beta}
=\left(\frac{d A_{0}}{d t}\right)^2 + 3\,H^2\,A_{0}^2\,,
\eeqa
\esubeqs
where the last step in \eqref{eq:DM-Q1} holds for
the RW metric \eqref{eq:RW-metric} and
the isotropic \textit{Ansatz} \eqref{eq:Ansatz-A}.

With the dimensionless variables \eqref{eq:dimensionless-variables},
the basic equations are given by the reduced
vector-field and FRW equations:
\bsubeqs\label{eq:system-v1-Dolgov-model}
\beqa\label{eq:system-v1-Dolgov-model-vdot}
\ddot{v} + 3\,h\,\dot{v} - 3\,h^2\,v &=& 0\,,
\\[2mm]\label{eq:system-v1-Dolgov-model-friedmann}
3\,h^2 &=& \lambda - \dot{v}^2 - 3\,h^2\,v^2\,\,,
\\[2mm]\label{eq:system-v1-Dolgov-model-hdot}
2\,\dot{h} + 3\,h^2 &=& \lambda - \dot{v}^2 - 3\,h^2\,v^2
-2\,\dot{h}\,v^2 - 4\,h\,v\,\dot{v} + 2\,\dot{v}^2\,,
\eeqa
\esubeqs
for $\lambda > 0$.
It is possible to set $\lambda=1$ by an appropriate
rescaling of $\tau$ and $1/h$, but we prefer to keep
$\lambda$ explicit, in order to facilitate comparison with
the models of App.~\ref{app:attractor-mathematics-ccp1-frw-model}
and Sec.~\ref{sec:Attractor-type-solutions}.
The system of differential equations can now be rewritten as
follows:
\bsubeqs\label{eq:system-v2-Dolgov-model}
\beqa
\ddot{v} + 3\,h\,\dot{v} - 3\,h^2\,v &=& 0\,,
\\[2mm]
3\,(1 + v^2)\,h^2 - \lambda + \dot{v}^2 &=& 0\,,
\\[2mm]
\big(1 + v^2\big)\,\dot{h} - \dot{v}^2 + 2\,h\,v\dot{v} &=& 0\,.
\eeqa
\esubeqs

Next, introduce new variables (using the natural logarithm `$\ln$'):
\bsubeqs\label{eq:new-variables-Dolgov-model}
\beqa
y_1 &\equiv& \dot{v}\,,
\quad
y_2 \equiv h\,v\,,
\\[2mm]
s &\equiv& \ln(a) - \ln(a_\text{start})\,,
\eeqa
\esubeqs
where $a=a(\tau)$ is the scale factor of the flat RW metric and
$a_\text{start}$ its value at $\tau=\tau_\text{start}$.
Writing \eqref{eq:system-v2-Dolgov-model} in terms of
the new variables \eqref{eq:new-variables-Dolgov-model} gives
the following \textit{autonomous} system of differential equations:%
\beqa\label{eq:odes-Dolgov-model}
y_1^{\prime} &=& F_1(y_1,\,y_2)\,,
\quad 
y_2^{\prime} = F_2(y_1,\,y_2)\,,
\eeqa
where the prime stands for differentiation with respect to
$s$ and
\bsubeqs\label{eq:F1-and-F2-Dolgov-model}
\beqa
F_1(y_1,\,y_2) &=& -3\,(y_1 - y_2)\,,
\\[2mm]
F_2(y_1,\,y_2) &=& \frac{y_1}{\lambda - y_1^2}\,
\Big(\lambda - y_1^2 + 3\,y_1\,y_2 - 6\,y_2^2\Big)\,.
\eeqa
\esubeqs
Recall that the system \eqref{eq:odes-Dolgov-model} is called autonomous
because the independent variable $s$ does not
occur explicitly (see, e.g., Refs.~\cite{Verhulst1996,Hahn1968}
for background material).

\subsection{Critical points}
\label{app-subsec:solution-Dolgov-model}

A critical point $(y_{10},y_{20})$ of system
\eqref{eq:odes-Dolgov-model} is defined as follows:
\beqa
F_1(y_1,\,y_2)\,\Big|_{y_{10},\,y_{20}} =
F_2(y_1,\,y_2)\,\Big|_{y_{10},\,y_{20}} = 0\,.
\eeqa
A straightforward calculation gives two such critical points,
\beqa\label{eq:special-critical-points-Dolgov-model}
y_{10}^{\pm} = \pm \,\sqrt{\lambda}\,/\,2\;,
\quad
y_{20}^{\pm} = \pm \,\sqrt{\lambda}\,/\,2\;,
\eeqa
corresponding to the asymptotic solutions
\beqa\label{eq:asymp-sol-Dolgov-model}
v_\text{asymp}^{\pm} = \pm\,\big(\sqrt{\lambda}\,/\,2\big)\;\tau,\,
\quad
h_\text{asymp} = \tau^{-1}\,,
\eeqa
in terms of the original
variables.\footnote{\label{ftn:deS-critical-point-Dolgov-model}There
is also a critical
point $(0,0)$, which corresponds
to de Sitter spacetime with $v=0$ and $h^2=\lambda/3$
if \eqref{eq:system-v1-Dolgov-model-friedmann}
is used as a constraint equation.
This critical point is not of interest to us now
and further discussion of this case will be omitted.
In addition, it can be shown that the critical
point $(0,0)$ is not asymptotically stable.}

\subsection{Stability analysis: Linearization}
\label{app-subsec:stability-analysis-linearization-Dolgov-model}

Make the following shift of variables:
\beqa
y_1 &=& y_{10} + Y_1\,,
\quad
y_2 = y_{20} + Y_2\,.
\eeqa
Then, \eqref{eq:odes-Dolgov-model} becomes
\bsubeqs\label{eq:odes-modified-Dolgov-model}
\beqa
\frac{dY_1}{d s} &=& -3\,Y_1 + 3\,Y_2
\equiv G_1\,,
\\[2mm]
\frac{dY_2}{d s} &=&
\frac{y_{10} + Y_1}{\lambda -\big(y_{10} + Y_1\big)^2}
\nonumber\\[1mm]&&
\times\Big(\lambda - \big(y_{10} + Y_1\big)^2 + 3\,\big(y_{10} + Y_1\big)
\,\big(y_{20} + Y_2\big) - 6\,\big(y_{20} + Y_2\big)^2\Big)
\equiv G_2\,.
\eeqa
\esubeqs

In order to prove that the critical points $(y_{10},y_{20})$
from \eqref {eq:special-critical-points-Dolgov-model}
are asymptotically stable solutions, it suffices to
consider small $Y_1$ and $Y_2$: $|Y_1| \ll |y_{10}|$ and
$|Y_2| \ll |y_{20}|$. We, then, find the following vector equation:
\bsubeqs\label{eq:linearized-system-Dolgov-model}
\beqa\label{eq:linearized-system-Dolgov-model-vector-eq}
\frac{d}{d s}\, Y(s)&=&A \cdot Y(s)\;+\;f(Y_1,\,Y_2)\,,
\eeqa
with the vectors
\beqa\label{eq:vectors-Dolgov-model}
Y(s)&=&\left(
  \begin{array}{c}
Y_1(s) \\
Y_2(s) \\
  \end{array}
\right) \,,
\qquad\quad
f(Y_1,\,Y_2)=\left(
  \begin{array}{c}
f_1(Y_1,\,Y_2) \\
f_2(Y_1,\,Y_2) \\
  \end{array}
\right)\,,
\eeqa
\esubeqs
and the constant matrix
\beqa\label{eq:matrices-Dolgov-model}
A&=&
\frac{1}{3}\left(
  \begin{array}{cc}
-9 & \;\;+9  \\
+1 & \;\;-9 \\
  \end{array}
\right)\,.
\eeqa

The eigenvalues of $A$ are both negative
($\sigma_1 = -2$, $\sigma_2 = -4$).
The vector component $f_1$ is zero
and $f_2$ is quadratic in $Y_1$ or $Y_2$ to leading order:
\bsubeqs
\beqa
\hspace*{-10mm}
f_1(Y_1,\,Y_2) &=& 0\,,
\\[2mm]
\hspace*{-10mm}
f_2(Y_1,\,Y_2) &=& \text{O}\,\big(
Y_1^2,\;Y_2^2,\;Y_1\,Y_2\big)\,,
\eeqa
so that the following bounds hold:
\beqa
\lim\limits_{Y_1,Y_2 \rightarrow 0}\;
\frac{f_1(Y_1,\,Y_2)}{\sqrt{Y_1^2 + Y_2^2}} =
\lim\limits_{Y_1,Y_2 \rightarrow 0}\;
\frac{f_2(Y_1,\,Y_2)}{\sqrt{Y_1^2 + Y_2^2}} = 0\,.
\eeqa
\esubeqs
With these results, the Poincar\'e--Lyapunov theorem
(Theorem 7.1 in Ref.~\cite{Verhulst1996};
see also Theorem 66.2 in Ref.~\cite{Hahn1968})
proves that the critical points $(y_{10},y_{20})$ from
\eqref{eq:special-critical-points-Dolgov-model}
are asymptotically stable (attractor) solutions.

\subsection{Stability analysis: Lyapunov function}
\label{app-subsec:stability-analysis-Lyapunov-function-Dolgov-model}

For completeness, we also give another proof which directly
starts from \eqref{eq:odes-modified-Dolgov-model}.
This proof relies on the construction of an appropriate Lyapunov
function $V$, in order to be able
to apply the second Lyapunov stability theorem~\cite{Verhulst1996,Hahn1968}.
The construction proceeds in three steps.

The first step is to
define the Lyapunov candidate function $V[s,\, Y_1,\, Y_2]$
with properties $V[s,0,0]=0$ and $V[s,\, Y_1,\, Y_2]>0$ for
$(Y_1,\, Y_2)\ne (0,0)$.
Specifically, take the following quadratic function:
\beq\label{eq:LtV-Ansatz}
V[s,\, Y_1,\, Y_2] = Y_1^2+ Y_2^2 +\big(Y_1- Y_2\big)^2\,,
\eeq
which has the required properties and no explicit dependence on $s$.

The second step is to calculate the orbital derivative,
\beq\label{eq:LtV-def}
L_{s}V\equiv
\frac{\partial V}{\partial s}
+\frac{\partial V}{\partial Y_1}\,G_1
+\frac{\partial V}{\partial Y_2}\,G_2\,,
\eeq
where the first derivative on the right-hand side
vanishes for the choice \eqref{eq:LtV-Ansatz}
and where $G_1$ and $G_2$ are defined by the right-hand sides
of \eqref{eq:odes-modified-Dolgov-model}.
The explicit result for $L_{s}V$ is
\beqa \label{eq:LtV-result}
L_{s}V &=&
-\,\frac{2}{4-(1+2\,Y_1)^2}\;
\Big[
19\,{\left( Y_{1} - Y_{2} \right) }^2
+ 8\,{Y_{2}}^2
+ 24\,\big({Y_{1}}^2\,Y_{2}- \,{Y_{1}}^3+ \,{Y_{2}}^3\big)
\nonumber\\[1mm]&&
- 28\,{Y_{1}}^4
+ 56\,{Y_{1}}^3\,Y_{2}
- 60\,{Y_{1}}^2\,{Y_{2}}^2
+ 48\,Y_{1}\,{Y_{2}}^3
\Big]
\,,
\eeqa
with Taylor expansion
\beqa \label{eq:LtV-result-series}
L_{s}V &=&
- \frac{2}{3}\;\Big[19\,\big(Y_1 - Y_2\big)^2 + 8\, Y_2^2\Big]
+\text{O}\big( Y^3 \big)\,.
\eeqa
The third and last step is to demonstrate that
\eqref{eq:LtV-result} implies the following inequality
for a sufficiently small domain of $Y_1$ and $Y_2$ around $Y_1=Y_2=0$:
\beq\label{eq:LtV-bound-zero-excluded}
L_{s}\, V\,\Big|_{(Y_1,\,Y_2)\ne(0,\,0)} < 0\,,
\eeq
where the strict inequality holds with the point $Y_1=Y_2=0$ excluded.
Result \eqref{eq:LtV-bound-zero-excluded}
implies that the function \eqref{eq:LtV-Ansatz} is
a genuine Lyapunov function.

The result \eqref{eq:LtV-bound-zero-excluded} for Lyapunov
function \eqref{eq:LtV-Ansatz} now establishes the fact
that the solution $Y(s) = 0$ is asymptotically stable
(Theorem 8.2 in Ref.~\cite{Verhulst1996} and
Theorem 25.2 in Ref.~\cite{Hahn1968}).

The precise mathematical definition of this asymptotic
attractor behavior can be found in, e..g, Sec.~5.2
of Ref.~\cite{Verhulst1996} (for a general discussion,
see also Sec.~35 of Ref.~\cite{Hahn1968}).
In short, arbitrary starting
values $\big(v_{0}(s_{0}),\,\dot{v}_{0}(s_{0})\big)$
in a sufficiently small  domain give a solution $\big(v(s),\,h(s)\big)$
which asymptotically approaches
the solution \eqref{eq:asymp-sol-Dolgov-model}
as `time' $s$ runs towards infinity.

\section{Attractor solution in a model with an $\boldsymbol{F_{A}}$ term}
\label{app:attractor-mathematics-ccp1-frw-model}

\subsection{ODEs and new variables}
\label{app-subsec:ODEs-and-new-variables-ccp1-frw-model}

In Ref.~\cite{EmelyanovKlinkhamer2011-CCP1-FRW},
we considered a single-vector-field model with the combination
$[\Lambda+\zeta_{0}\,(Q_3)^2+\kappa\,R^2 A^2]$
in the action density for $\Lambda> 0$.
For the case of $\zeta_{0} = 1$ and
$\kappa = - 1/2$, this corresponds to
having a vacuum-energy-density term $\epsilon=F_{A}$
in the action \eqref{eq:model-action}, with
$B_{\alpha}(x) \equiv 0$ and $F_{A}$ defined by \eqref{eq:def-FA-Q3A}.

The resulting dimensionless ODEs read:
\bsubeqs
\beqa
\hspace*{-8mm}&&
\ddot{v} + 3\,h\,\dot{v} - 6\,h^2\,v = 0\,,
\\[2mm]
\hspace*{-8mm}&&
3\,h^2 = \lambda - \big(\dot{v} + 3\,h\,v\big)^2 +
3\,h\,v\,\big(h\,v + 2\,\dot{v}\big) + r_M\,,
\\[2mm]
\hspace*{-8mm}&&
2\,\dot{h} + 3\,h^2 = \lambda - \big(\dot{v} + 3\,h\,v\big)^2
-\big[(4\,\dot{h} + 9\,h^2)\,v^2 - 2\,\dot{v}^2 - 2\,v\,\ddot{v} -
4h\,v\dot{v}\big] - \frac{1}{3}\,r_M\,,
\\[2mm]\label{eq:rM-ODE-ccp1-frw-model}
\hspace*{-8mm}&&
\dot{r}_M + 4\,h\,r_M = 0\,,
\eeqa
\esubeqs
for $\lambda > 0$.
With $h(\tau)\equiv \dot{a}(\tau)/a(\tau)$,
the solution of \eqref{eq:rM-ODE-ccp1-frw-model}
is known to be $r_M(\tau) \propto 1/a(\tau)^4$.
The system of differential equations can then be rewritten as follows:
\bsubeqs\label{eq:the-system}
\beqa
\ddot{v} + 3\,h\,\dot{v} - 6\,h^2\,v &=& 0\,,
\\[2mm]
3\,(1 + 2\,v^2)\,h^2 - \lambda + \dot{v}^2 -
r_{M\text{start}}\,\big(a_\text{start}/a\big)^4 &=& 0\,,
\\[2mm]
\big(1 + 2\,v^2\big)\,\dot{h} - \dot{v}^2 + 4\,h\,v\dot{v} +
\frac{2}{3}\,r_{M\text{start}}\,\big(a_\text{start}/a\big)^4 &=& 0\,,
\eeqa
\esubeqs
where $a=a(\tau)$ is the scale factor of the flat RW metric,
$a_\text{start}$ its value at $\tau=\tau_\text{start}$, and
$r_{M\text{start}}$ the corresponding starting value of $r_M$.

Next, introduce new variables:
\bsubeqs\label{eq:new-variables}
\beqa
y_1 &\equiv& \dot{v}\,,
\quad 
y_2 \equiv h\,v\,,
\\[2mm]
 s &\equiv& \ln(a) - \ln(a_\text{start})\,.
\eeqa
\esubeqs
Writing \eqref{eq:the-system} in terms of the
new variables \eqref{eq:new-variables} gives
the following \textit{nonautonomous} system of differential equations:
\beqa\label{eq:odes-plus}
y_1^{\prime} &=& F_1(s,\,y_1,\,y_2)\,,
\quad 
y_2^{\prime} = F_2(s,\,y_1,\,y_2)\,,
\eeqa
where the prime stands for differentiation with respect to $s$ and
\bsubeqs\label{eq:functions-F-and-G}
\beqa
F_1(s,\,y_1,\,y_2) &=& -3\,(y_1 - 2y_2)\,,
\\[2mm]
F_2(s,\,y_1,\,y_2) &=& y_1 + y_2\;
\frac{3\,y_1^2 - 12\,y_1\,y_2 - 2\,r_{M\text{start}}\;\exp[\,-4\,s\,]}
     {\lambda - y_1^2 + r_{M\text{start}}\;\exp[\,-4\,s\,]}\;\,.
\eeqa
\esubeqs

\subsection{Critical points}
\label{app-subsec:solution-ccp1-frw-model}

A critical point $(y_{10},y_{20})$ of system
\eqref{eq:odes-plus} is defined as follows:
\beqa
\lim_{s\to\infty}\,F_1(s,\,y_1,\,y_2)\,\Big|_{y_{10},\,y_{20}} =
\lim_{s\to\infty}\,F_2(s,\,y_1,\,y_2)\,\Big|_{y_{10},\,y_{20}} = 0\,.
\eeqa
A straightforward calculation gives two such critical points,
\beqa\label{eq:special_critical_points}
y_{10}^{\pm} = \pm \,\sqrt{2\,\lambda/5}\;,
\quad
y_{20}^{\pm} = \pm \,\sqrt{\lambda/10}\;,
\eeqa
corresponding to the asymptotic solutions
\beqa
v_\text{asymp}^{\pm} = \pm\,\sqrt{2\,\lambda/5}\;\,\tau,\,
\quad
h_\text{asymp} = (1/2) \,\tau^{-1}\,,
\eeqa
in terms of the original
variables.\footnote{\label{ftn:deS-critical-point-FA-model}%
It is obvious  that $(0,0)$ is also a critical point,
independent of the value of $r_{M\text{start}}$.
See Ftn.~\ref{ftn:deS-critical-point-Dolgov-model} for further comments.}

\subsection{Stability analysis}
\label{app-subsec:stability_analysis-ccp1-frw-model}

Make the following shift of variables:
\beqa
y_1 &=& y_{10} + Y_1\,,
\quad
y_2 = y_{20} + Y_2\,.
\eeqa
Then, \eqref{eq:odes-plus} becomes
\bsubeqs\label{eq:odes-modified-ccp1-frw-model}
\beqa
\frac{dY_1}{d s} &=& -3\,Y_1 + 6\,Y_2\,,
\\[2mm]
\frac{dY_2}{d s} &=& y_{10} + Y_1 +\big(y_{20} + Y_2\big)
\nonumber\\[1mm]&&
\times\;\frac{3\,(y_{10} + Y_1)^2 - 12\,(y_{10} + Y_1)
(y_{20} + Y_2) - 2\,r_{M\text{start}}\,\exp[\,-4\,s\,]}
{\lambda - (y_{10} + Y_1)^2 + r_{M\text{start}}\,\exp[\,-4\,s\,]}\;\,.
\eeqa
\esubeqs

In order to prove that the critical points $(y_{10},y_{20})$ from
\eqref{eq:special_critical_points} are asymptotically stable solutions,
it suffices to
consider small $Y_1$ and $Y_2$: $|Y_1| \ll |y_{10}|$ and
$|Y_2| \ll |y_{20}|$. We, then, find the following vector equation:
\bsubeqs\label{eq:linearized-system-ccp1-frw-model}
\beqa\label{eq:linearized-system-ccp1-frw-model-vector-eq}
\frac{d}{d s}\, Y(s)&=&A \cdot Y(s)\;+\;B(s) \cdot Y(s)\;
+\;f(s,\,Y_1,\,Y_2)\,,
\eeqa
with the vectors
\beqa\label{eq:vectors}
Y(s)&=&\left(
  \begin{array}{c}
Y_1(s) \\
Y_2(s) \\
  \end{array}
\right) \,,
\qquad\quad
f(s,\,Y_1,\,Y_2)=\left(
  \begin{array}{c}
f_1(s,\,Y_1,\,Y_2) \\
f_2(s,\,Y_1,\,Y_2) \\
  \end{array}
\right)\,,
\eeqa
and the matrices
\beqa\label{eq:matrices}
A&=&
\frac{1}{3}\left(
  \begin{array}{cc}
-9 & \;\;+18  \\
-1 & \;\;-18 \\
  \end{array}
\right)\,,
\qquad
B(s)\;=\;
\frac{20\,\alpha(s)}{3\,\big(3\,\lambda + 5\,\alpha(s)\big)}
\left(
  \begin{array}{cr}
0 & \;\;0  \\
1 & \;\;6 \\
  \end{array}
\right)\,,
\eeqa
in terms of the auxiliary variable
\beqa
\alpha(s) &\equiv& r_{M\text{start}}\,\exp[\,-4\,s\,]\,.
\eeqa
\esubeqs

The eigenvalues of $A$ are both negative
($\sigma_1 = -4$, $\sigma_2 = -5$) and
the matrix $B(s)$ vanishes as $s \rightarrow +\infty$.
The vector component $f_1$ is zero
and $f_2$ is quadratic in $Y_1$ or $Y_2$ to leading order,
\bsubeqs
\beqa
\hspace*{-10mm}
f_1(s,\,Y_1,\,Y_2) &=& 0\,,
\\[2mm]
\hspace*{-10mm}
f_2(s,\,Y_1,\,Y_2) &=& \text{O}\,\Bigg(
\frac{5\,\alpha(s) - 13\,
\lambda}{\big(5\,\alpha(s) + 3\,\lambda\big)^2}\;Y_1^2,\;
\frac{1}{5\,\alpha(s) + 3\,\lambda}\;Y_2^2,\;
\frac{25\,\alpha(s) + 27\,\lambda}
     {\big(5\,\alpha(s) + 3\,\lambda\big)^2}\;Y_1\,Y_2
\Bigg)\,,
\eeqa
so that the following bounds hold:
\beqa
\lim\limits_{Y_1,Y_2 \rightarrow 0}\;
\frac{f_1(s,\,Y_1,\,Y_2)}{\sqrt{Y_1^2 + Y_2^2}} =
\lim\limits_{Y_1,Y_2 \rightarrow 0}\;
\frac{f_2(s,\,Y_1,\,Y_2)}{\sqrt{Y_1^2 + Y_2^2}} = 0\,.
\eeqa
\esubeqs
With these results, the Poincar\'e--Lyapunov theorem
(Theorem 7.1 in Ref.~\cite{Verhulst1996};
see also the discussion below Theorem 66.2 in Ref.~\cite{Hahn1968})
proves that the critical points
$(y_{10},y_{20})$ from \eqref{eq:special_critical_points}
are asymptotically stable (attractor) solutions.
The above discussion provides the details for the result announced in the
Note Added of Ref.~\cite{EmelyanovKlinkhamer2011-CCP1-FRW}.

\end{appendix}


\end{document}